\documentclass[a4paper,12pt]{article}
\pdfoutput=1
\usepackage{cite}
\usepackage{amssymb}
\usepackage{tipa}
\usepackage{mathrsfs}
\usepackage{bbm}
\usepackage{epsf,amsmath}
\usepackage{graphicx,subfigure}
\usepackage{colortbl}
\usepackage{booktabs,multirow,makecell}
\usepackage{float}
\usepackage{threeparttable}
\usepackage{subfigure}
\usepackage{color}

\newlength{\dinwidth}
\newlength{\dinmargin}
\def\be{\begin{equation}}
\def\ee{\end{equation}}
\def\ba{\begin{eqnarray}}
\def\ea{\end{eqnarray}}

\newcommand{\half}{\frac{1}{2}}
\allowdisplaybreaks

\begin{document}

\title{\bf Study of the weak annihilation contributions\\ in charmless \boldmath{$B_s\to VV$} decays}

\author{Qin Chang$^{a,b}$, Xiaonan Li$^{a}$, Xin-Qiang Li$^{b}$ and Junfeng Sun$^{a}$\\
{ $^a$\small Institute of Particle and Nuclear Physics, Henan Normal University, Henan 453007, China}\\
{ $^b$\small Institute of Particle Physics and Key Laboratory of Quark and Lepton Physics~(MOE),} \\
{     \small Central China Normal University, Wuhan, Hubei 430079, China}}

\date{}
\maketitle

\begin{abstract}
\noindent In this paper, in order to probe the spectator-scattering and weak annihilation contributions in charmless $B_s\to VV$ (where $V$ stands for a light vector meson) decays, we perform the $\chi^2$-analyses for the end-point parameters within the QCD factorization framework, under the constraints from the measured $\bar B_{s}\to$$\rho^0\phi$, $\phi K^{*0}$, $\phi \phi$ and $K^{*0}\bar K^{*0}$ decays. The fitted results indicate that the end-point parameters in the factorizable and nonfactorizable annihilation topologies are non-universal, which is also favored by the charmless $B\to PP$ and $PV$ (where $P$ stands for a light pseudo-scalar meson) decays observed in the previous work. Moreover, the abnormal polarization fractions $f_{L,\bot}(\bar B_{s}\to K^{*0}\bar K^{*0})=(20.1\pm7.0)\%\,,(58.4\pm8.5)\%$ measured by the LHCb collaboration can be reconciled through the weak annihilation corrections. However, the branching ratio of $\bar B_{s}\to\phi K^{*0}$ decay exhibits a tension between the data and theoretical result, which dominates the contributions to $\chi_{\rm min}^2$ in the fits. Using the fitted end-point parameters, we update the theoretical results for the charmless $B_s\to VV$ decays, which will be further tested by the LHCb and Belle-II experiments in the near future.
\end{abstract}

\newpage

\section{Introduction}

Non-leptonic $B$-meson weak decays play an important role in testing the flavor dynamics of the Standard Model~(SM) and exploring possible hints of New Physics beyond it. Theoretically, one of the main obstacles for a reliable prediction on these decays is how to evaluate precisely the hadronic matrix elements of local operators between the initial and final hadronic states, especially due to the nontrivial QCD dynamics involved. To this end, several attractive QCD-inspired approaches, such as QCD factorization~(QCDF)~\cite{Beneke:1999br,Beneke:2000ry}, perturbative QCD~(pQCD)~\cite{Keum:2000ph,Keum:2000wi} and soft-collinear effective theory~(SCET)~\cite{Bauer:2000yr,Bauer:2001yt,Beneke:2002ph,Beneke:2002ni}, have been proposed in the last decades. However, the convolution integrals of the parton-level hard kernels with the asymptotic forms of light-cone distribution amplitudes (LCDAs) generally suffer from the end-point divergence in the weak annihilation~(WA) amplitudes. This divergence limits the predictive power and introduces large theoretical uncertainties.

In the QCDF approach, the end-point divergent integrals, signalling of infrared-sensitive contributions, are usually parameterized by two complex quantities $X_{A}$ and $X_{L}$ that are defined, respectively, by~\cite{Beneke:2003zv,du2}
\begin{eqnarray}
\label{Xa}
\int_{0}^{1}\frac{dx}{x}&\to& X_{A}(\rho_A,\phi_A)=(1+\rho_{A}e^{i\phi_{A}})\ln\frac{m_{b}}{\Lambda_{h}}\,,\\
\label{XL}
\int_{0}^{1}\frac{dx}{x^2}&\to& X_{L}(\rho_A,\phi_A)=(1+\rho_{A}e^{i\phi_{A}})\frac{m_{b}}{\Lambda_h}\,,
\end{eqnarray}
where $\Lambda_{h}=0.5~{\rm GeV}$, and the two phenomenological parameters $\rho_A$ and $\phi_A$ account for the strength and possible strong phase of WA contributions near the end-point, respectively. In addition, the spectator-scattering amplitudes also involve the end-point divergence, which is dealt with the same manner by introducing the complex quantity  $X_{H}(\rho_H,\phi_H)=X_{A}(\rho_A,\phi_A)|_{A\to H}$. The numerical values of $\rho_{A\,,H}$ and $\phi_{A\,,H}$ are unknown and can only be inferred by fitting them to the experimental data so far. While weakening the predictive power of the QCDF approach, the parameterization scheme provides a feasible way to explore the WA effects from a phenomenological point of view.

Theoretically, the WA contributions with possible strong phase have attracted a lot of attention in the past few years, for instance, in Refs.~\cite{Arnesen:2006dc,Xiao:2011tx,Ali:2007ff,Chang:2012xv,Li:2015xna,
Huang:2005if,Li:2004ep,Yang:2005nra,Cheng:2009mu,Cheng:2009cn,Cheng:2008gxa,
Bobeth:2014rra,Zhu:2011mm,Wang:2013fya,Chang:2014rla,Chang:2014yma,Sun:2014tfa,
Chang:2015wba,Chang:2016ren,Cheng:2014rfa}. Traditionally, both $\rho_A$ and $\phi_A$ are treated as universal parameters for different kinds of annihilation topologies. However, a global fit for the end-point parameters indicates that, while a relatively large end-point parameter is needed for the decays related by isospin symmetry, there exist some tensions in $B\to \phi K^\ast$ and $B\to \pi K$ decays~\cite{Bobeth:2014rra}, with the latter exhibiting the so-called ``$\pi K$ CP-asymmetry puzzle''.\footnote{The direct CP asymmetries $A_{CP}^{\pi^0 K^+}$ in $B^{\pm}\to \pi^0 K^{\pm}$ and $A_{CP}^{\pi^- K^+}$ in $B^0(\bar B^0)\to \pi^{\mp} K^{\pm}$ decays are expected to be roughly the same~\cite{Bell:2015koa,Fleischer:2007mq,Liu:2015upa}. However, the current experimental data show a significant difference between them, $A_{CP}^{\pi^0 K^+}-A_{CP}^{\pi^- K^+}=0.122\pm0.022$~\cite{Amhis:2014hma,Olive:2016xmw}, deviating from zero at about $5.6~\sigma$ level.} In Refs.~\cite{Zhu:2011mm,Wang:2013fya}, after studying carefully the flavor dependence of the end-point parameters in charmless $B\to PP$ (where $P$ stands for a light pseudo-scalar meson) decays, the authors suggest that the end-point parameters should be topology-dependent. Such a topology-dependent parameterization scheme is also favored by most of the charmless $B\to PP$ and $PV$ (where $V$ stands for a light vector meson) decays, as demonstrated in Refs.~\cite{Chang:2014yma,Sun:2014tfa,Chang:2015wba}, and it could provide a possible solution to the well-known ``$\pi K$ CP-asymmetry puzzle''~\cite{Chang:2014rla}. In addition, using the recent measurements of the pure annihilation $B_s\to\pi^+\pi^-$ and $B_d\to K^+K^-$ decays by the CDF~\cite{CDFanni}, Belle~\cite{Duh:2012ie} and LHCb collaborations~\cite{LHCbanni,LHCbanni16}, the authors of Refs.~\cite{Zhu:2011mm,Wang:2013fya} find significant flavor-symmetry breaking effects in the nonfactorizable annihilation contributions.

Experimentally, due to the rapid development of dedicated heavy-flavor experiments, more precise measurements of non-leptonic $B$ decays will be available. As reported in Ref.~\cite{Gershon:2014rga}, for instance, over $10^{11}$ $b\bar{b}$ quark pairs are produced per $\mathrm{fb}^{-1}$ of data at the LHCb experiment. Furthermore, after the high-luminosity upgrade, a dataset of $50~\mathrm{fb}^{-1}$ will be accumulated~\cite{Aaij:2010gn,Bediaga:2012py,Aaij:2014jba,LHCb:upgrade}. In addition, most recently, the SuperKEKB/Belle-II experiment has started test operations and succeeded in circulating and storing beams in the electron and positron rings. The annual integrated luminosity is expected to reach up to $13~\mathrm{ab}^{-1}$, and over $10^{10}$ samples of $b\bar{b}$ quark pairs will be accumulated by the Belle-II experiment~\cite{Abe:2010gxa}. Thus, the forthcoming measurements of not only $B_{u,d}$ but also $B_s$ decays are expected to reach a high accuracy, which could provide us with a clearer picture  of the WA contributions in these decays.

In this paper, motivated by the recent theoretical studies and the bright experimental prospects, we will investigate the WA contributions in charmless $B_s\to VV$ decays, which involve more observables than in charmless $B\to PP$ and $PV$ decays and may provide much stronger constraints on the end-point parameters.  In addition, using the obtained values of the end-point parameters, we will update the theoretical results for charmless $B_s\to VV$ decays within the QCDF framework.

Our paper is organized as follows. In section 2, we review briefly the WA amplitudes within the QCDF framework and observables for charmless $B_s\to VV$ decays. Section 3 is devoted to the numerical results and discussions. Finally, we give our conclusion in section 4.

\section{Brief review of the theoretical framework for charmless \boldmath{$B_s\to VV$} decays}

\subsection{Amplitudes in QCD factorization}

In the SM, the effective weak Hamiltonian for non-leptonic $B$-meson decays is given by~\cite{Buchalla:1995vs,Buras:1998raa}
\begin{align} \label{Heff}
\mathcal{H}_\mathrm{eff} &= \frac{G_{F}}{\sqrt{2}}\,\bigg\{V_{ub}V_{up}^{*}\left(C_{1}O_{1}^{u} +C_{2}O_{2}^{u}\right) +V_{cb}V_{cp}^{*}\left(C_{1}O_{1}^{c}+C_{2}O_{2}^{c}\right)\nonumber\\
& \hspace{1.2cm} -V_{tb}V_{tp}^{*}\Big(\sum_{i=3}^{10}C_{i}O_{i}+C_{7\gamma}O_{7\gamma} +C_{8g}O_{8g}\Big)\bigg\}+\text{h.c.}\,,
\end{align}
where $V_{qb}V_{qp}^{*}$, with $q\in\{u,c,t\}$ and $p\in\{d,s\}$, are products of the Cabibbo-Kobayashi-Maskawa (CKM) matrix elements, and $C_{i}$ the Wilson coefficients of the effective operators $O_{i}$. Starting with the effective Hamiltonian and following the strategy proposed in Ref.~\cite{Lepage:1980fj}, Beneke \textit{et al.} proposed the QCDF approach to evaluate the hadronic matrix elements~\cite{Beneke:1999br,Beneke:2000ry}, which is now being widely used to analyze the $B$-meson weak decays~(see, for instance, Refs.~\cite{Beneke:2003zv,Bell:2015koa,Beneke:2005we,Huber:2016xod,du01,du02,Beneke:2006hg,Beneke:2009ek,Chang:2016eto}). The theoretical framework for charmless $B\to VV$ decays has also been fully developed (cf. Refs.~\cite{Beneke:2006hg,Cheng:2008gxa,Cheng:2009mu} for details). In this paper, we follow the same conventions as in Refs.~\cite{Beneke:2006hg,XQLi}.

Within the QCDF framework, after performing the convolution integrals of the $\mathcal{O}(\alpha_s)$ hard kernels with the asymptotic forms of the light-meson LCDAs, one gets the following basic building blocks of the WA amplitudes in charmless $B\to VV$ decays~\cite{Beneke:2006hg,XQLi}:
\begin{align}\label{a1iz}
A_{1}^{i,0} &\simeq A_{2}^{i,0} \simeq 18\pi\alpha_{s}\left[\left(X_{A}^{i}-4+\frac{\pi^2}{3}\right)+r_{\chi}^{V_{1}} r_{\chi}^{V_{2}}\left(X_{A}^{i}-2\right)^2\right]\,,\\[0.1cm]
A_{3}^{i,0} &\simeq
18\pi\alpha_{s}\left(r_{\chi}^{V_{1}}-r_{\chi}^{V_{2}}\right)\left[-(X_{A}^{i})^2 +2X_{A}^{i}-4+\frac{\pi^2}{3}\right]\,,\label{a3iz} \\[0.1cm]
A_{3}^{f,0} &\simeq
18\pi\alpha_{s}\left(r_{\chi}^{V_{1}}+r_{\chi}^{V_{2}}\right)\left(2X_{A}^{f}-1\right)
\left(2-X_{A}^{f}\right)\,,
\end{align}
for the non-vanishing longitudinal contributions, and
\begin{align}
A_{1}^{i,+} &\simeq A_{2}^{i,+} \simeq 18\pi\alpha_{s}\frac{m_{1}m_{2}}{m_{B}^2}\left[2(X_{A}^{i})^2-3X_{A}^{i}+6 -\frac{2\pi^2}{3}\right]\,,\\[0.1cm]
A_{1}^{i,-} &\simeq A_{2}^{i,-}\simeq
18\pi\alpha_{s}\frac{m_{1}m_{2}}{m_{B}^2}\left(\half X_{L}^{i}+\frac{5}{2}-\frac{\pi^2}{3}\right),\\[0.1cm]
A_{3}^{i,-} &\simeq
18\pi\alpha_{s}\left(\frac{m_{1}}{m_{2}}r_{\chi}^{V_{2}}-\frac{m_{2}}{m_{1}} r_{\chi}^{V_{1}}\right)\left[(X_{A}^{i})^2-2X_{A}^{i}+2\right]\,,\label{a3im} \\[0.1cm]
A_{3}^{f,-} &\simeq
18\pi\alpha_{s}\left(\frac{m_{1}}{m_{2}}r_{\chi}^{V_{2}}+\frac{m_{2}}{m_{1}} r_{\chi}^{V_{1}}\right)\left[2(X_{A}^{f})^2-5X_{A}^{f}+3\right]\,,\label{a3fm}
\end{align}
for the transverse ones, where the superscripts $0\,,\pm$ refer to the vector-meson helicities. Here $r_{\chi}^{V}=\frac{2m_Vf_V^{\bot}}{m_bf_V}$, with $f_V$ and $f_V^{\bot}$ denoting the longitudinal and transverse vector-meson decay constants; $m_B$ and $m_{1,2}$ are the masses of the initial and final states, respectively.

The two complex quantities $X_{A}$ and $X_{L}$ are introduced in Eqs.~\eqref{a1iz}--\eqref{a3fm} to parameterize the end-point divergence (cf. Eqs.~(\ref{Xa}) and (\ref{XL})).
In addition, we have distinguished the WA contributions with the gluon emitted either from the initial (marked by the superscript ``$i$") or from the final state (marked by the superscript ``$f$"), corresponding to the nonfactorizable and the factorizable annihilation topologies, respectively. For the factorizable annihilation topologies, as argued in Refs.~\cite{Zhu:2011mm,Wang:2013fya}, since all decay constants have been factorized out from the hadronic matrix elements, the building blocks $A_{3}^{f,0}$ and $A_{3}^{f,-}$ are independent of the initial states.
However, for the nonfactorizable annihilation topologies, $X_{A,L}^i$ are generally non-universal for $B_{u,d}$ and $B_s$ decays~\cite{Zhu:2011mm,Wang:2013fya}.
Besides, an additional complex quantity, $X_{H}(\rho_H,\phi_H)=X_{A}(\rho_A,\phi_A)|_{A\to H}$,  is introduced to parameterize the end-point divergence in the hard spectator-scattering~(HSS) amplitudes~(cf. Refs.~\cite{Beneke:2003zv,Beneke:2006hg} for details).

With the above prescriptions for the WA amplitudes, we consider in this paper the following $B_{s}\to VV$ decay modes:\footnote{The expressions for the decay amplitudes given below should be multiplied with $V_{pb}V_{pd}^{\ast}$ (for $\Delta D=1$ transition) and $V_{pb}V_{ps}^{\ast}$ (for $\Delta S=1$ transition) and summed over $p=u,c$.}
\begin{enumerate}
\item[(i)] $\Delta D=1$ transition: the color-allowed tree-dominated $\bar B_{s}\to \rho^- K^{*+}$, the color-suppressed tree-dominated $\bar B_{s}\to \rho^0 K^{*0}$ and $\omega K^{*0}$, as well as the penguin-dominated $\bar B_{s}\to \phi K^{*0}$ decay, the amplitudes of which are given, respectively, as~\cite{Beneke:2003zv,Beneke:2006hg}
\begin{align}
\label{eq:Arhomkp}
{\cal{A}}_{\bar B_{s}\to \rho^{-}K^{*+}}^{h}&=A_{K^{*}\rho}^{h}
\left[\delta_{pu}\alpha_{1}^{p,h}+\alpha_{4}^{p,h}+\alpha_{4,\mathrm{EW}}^{p,h} +\beta_{3}^{p,h}-\half\beta_{3,\mathrm{EW}}^{p,h}\right],\\[0.1cm]
\label{eq:Arho0k0}
\sqrt{2}{\cal{A}}_{\bar B_{s}\to \rho^{0}K^{*0}}^{h}&=A_{K^{*}\rho}^{h}
\left[\delta_{pu}\alpha_{2}^{p,h}-\alpha_{4}^{p,h}+\frac{3}{2}\alpha_{3,\mathrm{EW}}^{p,h} +\frac{1}{2}\alpha_{4,\mathrm{EW}}^{p,h}
-\beta_{3}^{p,h}+\frac{1}{2}\beta_{3,\mathrm{EW}}^{p,h}\right],\\[0.1cm]
\label{eq:Awk0}
\sqrt{2}{\cal{A}}_{\bar B_{s}\to \omega K^{*0}}^{h}&=A_{K^{*}\omega}^{h}
\left[\delta_{pu}\alpha_{2}^{p,h}+2\alpha_{3}^{p,h}+\alpha_{4}^{p,h} +\frac{1}{2}\alpha_{3,\mathrm{EW}}^{p,h}-\frac{1}{2}\alpha_{4,\mathrm{EW}}^{p,h}+\beta_{3}^{p,h} -\frac{1}{2}\beta_{3,\mathrm{EW}}^{p,h}\right],\\[0.1cm]
\label{eq:Aphik}
{\cal{A}}_{\bar B_{s}\to \phi K^{*0}}^{h}&=A_{K^{*}\phi}^{h}
\left[\alpha_{3}^{p,h}-\half\alpha_{3,\mathrm{EW}}^{p,h}\right] +
A_{\phi  K^{*}}^{h}\left[\alpha_{4}^{p,h}-\half\alpha_{4,\mathrm{EW}}^{p,h}+\beta_{3}^{p,h}- \half\beta_{3,\mathrm{EW}}^{p,h}\right]\,.
\end{align}

\item[(ii)] $\Delta S=1$ transition: the penguin-dominated $\bar B_{s}\to K^{*+} K^{*-}$, $K^{*0} \bar K^{*0}$, and $\phi \phi$, $\rho^0 \phi$, $\omega \phi$, as well as the pure annihilation $\bar B_{s}\to \rho^{+} \rho^{-}$, $\rho^{0} \rho^{0}$, $\rho \omega$, $\omega \omega$ decays, the amplitudes of which are given, respectively, as~\cite{Beneke:2003zv,Beneke:2006hg}
\begin{align}
{\cal{A}}_{\bar B_{s}\to \bar K^{*-}K^{*+}}^{h}&=A_{K^{*}\bar K^{*}}^{h}
\left[\delta_{pu}\alpha_{1}^{p,h}+\alpha_{4}^{p,h}+\alpha_{4,\mathrm{EW}}^{p,h}\right]
\nonumber\\
&+B_{\bar K^{*}K^{*}}^{h}
\left[\delta_{pu}b_{1}^{p,h}+b_{3}^{p,h}+2b_{4}^{p,h}-\half b_{3,\mathrm{EW}}^{p,h}+\half b_{4,\mathrm{EW}}^{p,h}\right],\label{KpKm}\\[0.1cm]
\label{KK}
{\cal{A}}_{\bar B_{s}\to \bar K^{*0}K^{*0}}^{h}&=A_{K^{*}\bar K^{*}}^{h}
\left[\alpha_{4}^{p,h}-\half\alpha_{4,\mathrm{EW}}^{p,h}\right]+B_{\bar K^{*}K^{*}}^{h}
\left[b_{3}^{p,h}+2b_{4}^{p,h}-\half b_{3,\mathrm{EW}}^{p,h}-b_{4,\mathrm{EW}}^{p,h}\right],\\[0.1cm]
\label{eq:Aphiphi}
\half{\cal{A}}_{\bar B_{s}\to \phi\phi}^{h}&=A_{\phi\phi}^{h}
\left[\alpha_{3}^{p,h}+\alpha_{4}^{p,h}-\half\alpha_{3,\mathrm{EW}}^{p,h} -\half\alpha_{4,\mathrm{EW}}^{p,h} +\beta_{3}^{p,h}+\beta_{4}^{p,h}-\half\beta_{3,\mathrm{EW}}^{p,h} -\half\beta_{4,\mathrm{EW}}^{p,h}\right],\\[0.1cm]
\label{eq:Arho0phi}
\sqrt{2}{\cal{A}}_{\bar B_{s}\to \rho^{0}\phi}^{h}&=A_{\phi\rho}^{h}
\left[\delta_{pu}\alpha_{2}^{p,h}+\frac{3}{2}\alpha_{3,\mathrm{EW}}^{p,h}\right],\\[0.1cm]
\label{wphi}
{\cal{A}}_{\bar B_{s}\to \omega\phi}^{h}&=\sqrt{2}A_{\phi\omega}^{h}
\left[\delta_{pu}\alpha_{2}^{p,h}+2\alpha_{3}^{p,h} +\frac{1}{2}\alpha_{3,\mathrm{EW}}^{p,h}\right],\\[0.1cm]
\label{eq:Arho0rho0}
{\cal{A}}_{\bar B_{s}\to \rho^{0}\rho^{0}}^{h}&=B_{\rho\rho}^{h}
\left[\delta_{pu}b_{1}^{p,h}+2b_{4}^{p,h} +\frac{1}{2}b_{4,\mathrm{EW}}^{p,h}\right],\\[0.1cm]
\label{eq:Arhoprhom}
{\cal{A}}_{\bar B_{s}\to \rho^{+}\rho^{-}}^{h}&=B_{\rho^-\rho^+}^{h}
\left[\delta_{pu}b_{1}^{p,h}+b_{4}^{p,h}+b_{4,\mathrm{EW}}^{p,h}\right]+
B_{\rho^+\rho^-}^{h}
\left[b_{4}^{p,h}-\frac{1}{2}b_{4,\mathrm{EW}}^{p,h}\right],\\[0.1cm]
\label{eq:Arho0w}
{\cal{A}}_{\bar B_{s}\to \rho^{0}\omega}^{h}&=B_{\rho\omega}^{h}
\left[\delta_{pu}b_{1}^{p,h}+\frac{3}{2}b_{4,\mathrm{EW}}^{p,h}\right],\\[0.1cm]
\label{eq:Aww}
{\cal{A}}_{\bar B_{s}\to \omega\omega}^{h}&=B_{\omega\omega}^{h}
\left[\delta_{pu}b_{1}^{p,h}+2b_{4}^{p,h}+\frac{1}{2}b_{4,\mathrm{EW}}^{p,h}\right].
\end{align}
\end{enumerate}
In the above decay amplitudes, the vertex, penguin and spectator-scattering corrections are encoded in the effective coefficients $\alpha_i^{p,h}$ (cf. Ref.~\cite{Beneke:2003zv,Beneke:2006hg} for details), and the WA contributions are denoted by $\beta_i^{p,h}$ (or $b_i^{p,h}$), which are defined, respectively, by~\cite{Beneke:2003zv,Beneke:2006hg}
\begin{align}
{\beta}_{i}^{p,h} &= b_{i}^{p,h} B^h_{M_{1}M_{2}}/A^h_{M_{1}M_{2}} \label{betai}, \\[0.1cm]
b_{1}^h &= \frac{C_{F}}{N_{c}^{2}}\, C_{1} A_{1}^{i,h},
   \quad \quad
b_{2}^h = \frac{C_{F}}{N_{c}^{2}}\, C_{2} A_{1}^{i,h}
   \label{b12}, \\[0.1cm]
b_{3}^{p,h} &= \frac{C_{F}}{N_{c}^{2}}\,
   \left[ C_{3} A_{1}^{i,h} + C_{5}( A_{3}^{i,h} + A_{3}^{f,h} )
   +N_{c} C_{6} A_{3}^{f,h} \right]\label{b3}, \\[0.1cm]
b_{4}^{p,h} &=
   \frac{C_{F}}{N_{c}^{2}}\,
   \left[ C_{4} A_{1}^{i,h} + C_6 A_2^{i,h} \right]
   \label{b4}, \\[0.1cm]
b_{3,\rm EW}^{p,h} &=
   \frac{C_{F}}{N_{c}^{2}}\,
   \left[ C_9 A_1^{i,h} + C_7 ( A_3^{i,h} + A_3^{f,h} )
   + N_c C_8 A_3^{f,h} \right]
   \label{b3ew}, \\[0.1cm]
b_{4,\rm EW}^{p,h} &=
   \frac{C_{F}}{N_{c}^{2}}\,
   \left[ C_{10} A_1^{i,h} + C_8 A_2^{i,h} \right]\,
   \label{b4ew}.
   \end{align}

Based on the previous studies~\cite{Beneke:2006hg,Bobeth:2014rra} and the amplitudes given above, $B_s\to VV$ decay modes can  be classified as follows according to their sensitivities to the WA and/or HSS corrections:
\begin{itemize}
\item The pure annihilation decay modes: Because both WA and HSS corrections involve the undetermined end-point contributions, the interference between them presents an obstacle for precisely probing the WA contributions from data.  Fortunately, such problem can be avoided by using the pure annihilation decay modes, which can be easily seen from Eqs.~\eqref{eq:Arho0rho0}, \eqref{eq:Arhoprhom}, \eqref{eq:Arho0w} and \eqref{eq:Aww}. In this paper, the $\bar B_{s}\to \rho^{0}\rho^{0}\,,\rho^{+}\rho^{-}\,,\rho^{0}\omega$ and $\omega\omega$ decays belong to this category, but unfortunately, they have not been measured for now.
\item The color-suppressed tree- and electroweak or QCD flavor-singlet penguin-dominated decays: Only two decay modes fall into this category,  namely, $\bar B_{s}\to \rho^{0}\phi$ and $\omega\phi$. From Eqs.~\eqref{eq:Arho0phi}  and \eqref{wphi},  one can find that these decays are characterized by an interplay of the color- and CKM-suppressed tree amplitude,  $\alpha_2$, electroweak penguin amplitude, $\alpha_{3,{\rm EW}}^{c}$, and flavor-singlet QCD penguin amplitude, $\alpha_{3}^{c}$. For $\bar B_{s}\to\omega\phi$ decay, due to a partial cancellation between the QCD and electroweak penguin contributions, the largest partial amplitude is $\alpha_2$~\cite{Beneke:2006hg}, which is very sensitive to the HSS corrections. For $\bar B_{s}\to \rho^{0}\phi$ decay, $\alpha_2$ is also nontrivial even though it is numerically smaller than $\alpha_{3,{\rm EW}}^{c}$ when $\rho_H$ is small. More importantly, a remarkable feature of such two decays is that their amplitudes are irrelevant to the WA contributions and, therefore, very suitable for probing the HSS corrections. Recently, the branching ratio of $\bar B_{s}\to \rho^{0}\phi$ decay has been measured by the LHCb collaboration with a statistical significance of about $4\sigma$~\cite{Aaij:2016qnm}.
\item The color-suppressed tree-dominated  $\Delta D=1$ decays: This class includes $\bar B_{s}\to \rho^{0}K^{*0}$ and  $\omega K^{*0}$ decays, whose amplitudes are given by Eqs.~\eqref{eq:Arho0k0} and \eqref{eq:Awk0}, respectively. The CKM-factors relevant to the effective coefficients in their amplitudes are at the same order, $\sim\lambda^3$, and therefore one can roughly find that $\alpha_2$ dominates their amplitudes. Considering further the fact that the HSS contribution in $\alpha_2$ is proportional to the largest Wilson coefficient $C_1$, we can generally expect that these decays present strong constraints on the HSS end-point parameters even though they are not as ``clean" as $\bar B_{s}\to \rho^{0}\phi$ decay due to the interference induced by $\beta_3^c$. However, there is no available data for  these decays for now.
\item The QCD penguin-dominated decays:  This class contains the residual decays,  except for $\bar B_{s}\to \rho^{-}K^{*+}$, considered in this paper, among which $\bar B_{s}\to \phi K^{*0}$, $\phi \phi$ and $K^{*0} \bar K^{*0}$ decays have been measured. From their amplitudes, Eqs.~\eqref{eq:Aphik},  \eqref{KK} and \eqref{eq:Aphiphi}, one can find that the effective color-allowed QCD penguin amplitude, $\hat{\alpha}_{4}^c\equiv \alpha_{4}^c+\beta_3^c$, plays a dominant role~\cite{Beneke:2006hg}; and the penguin-annihilation amplitude, $\beta_4^c$, presents the first subdominant contribution besides $\hat{\alpha}_{4}^c$ for the longitudinal amplitude of the last two decay modes~\cite{Beneke:2006hg}. Considering further the fact that the HSS contribution in $\alpha_{4}^c$ is trivial compared to the LO contribution $C_4+C_3/{N_C}$,  we can generally conclude that such three QCD penguin-dominated decays are suitable for probing the WA contribution from data even through they are not as ``clean'' as the pure annihilation decay modes.\\
    It should be noted that, such an expectation or such a conclusion is valid  only when $\rho_H$ is not very large especially for  the decays involving $\alpha_3^c$, for instance,  $\bar B_{s}\to \phi K^{*0}$ and $\phi \phi$ decays. The HSS correction in $\alpha_3^c$ with a large $\rho_H$ can bring about a significant correction.
\item The color-allowed tree-dominated  $\Delta D=1$ decay: In this paper, only $\bar B_{s}\to \rho^{-}K^{*+}$ decay belongs to this class. For this decay mode, the effects of HSS and WA contributions are generally not very significant,  because of the dominant role played by the color-allowed tree amplitude, $\alpha_1$. It also has not been measured.
\end{itemize}

\subsection{Observables}

Using the amplitudes given above, the observables for $B_s\to VV$ decays can be defined as follows. The most important observables are the CP-averaged branching ratio and direct CP asymmetry, which are defined theoretically as~\cite{Bobeth:2014rra}
\begin{eqnarray}
\label{eq:br}
{\cal B}[B_{s}\to f]&=&\frac{\tau_{B_s}}{2}(\Gamma[\bar{B}_{s}\to \bar{f}]+\Gamma[B_{s}\to f])\,,\\
\label{eq:cp}
{\cal{A}}_{CP}&=&\frac{\Gamma[\bar{B}_{s}\to \bar{f}]-\Gamma[B_{s}\to f]}{\Gamma[\bar{B}_{s}\to \bar{f}]+\Gamma[B_{s}\to f]}\,,
\end{eqnarray}
respectively. The decay rates should be summed over the polarization state~($h^{\prime}=L,\parallel,\perp$) for evaluating the ``whole" observables.  Following the convention of Ref.~\cite{Beneke:2006hg}, the polarization amplitudes can be easily obtained from the helicity amplitudes through the relations, $\bar{A}_L=\bar{A}_0$ and $\bar{A}_{\parallel,\perp}=\frac{\bar{A}_-\pm \bar{A}_+}{\sqrt{2}}$.

Besides branching ratio and CP asymmetry given by Eqs.~\eqref{eq:br} and \eqref{eq:cp},  the two-body $B_s\to VV$ decays with cascading decays $V\to PP$ provide additional observables in the full angular analysis of the 4-body final state~\cite{Korner:1979ci}.  There are polarization fractions and relative phases defined by
\begin{eqnarray}
\label{eq:angobs}
f_{h^{\prime}}^{\bar{B}_s}=\frac{|\bar{A}_{h^{\prime}}|^2}{\sum_{h^{''}}|
\bar{A}_{h^{''}}|^2}\,,\quad
\phi_{\parallel,\perp}^{\bar{B}_s}={\rm arg} \frac{\bar{A}_{\parallel,\perp}}{\bar{A}_{L}}\,,
\end{eqnarray}
for $\bar{B}_s$ decays. The same quantities for $B_s$ decays are obtained by the replacement $\bar{A}_{h^{\prime}}\to A_{h^{\prime}}$.  The CP-averaged polarization fractions  and  CP asymmetries are given, respectively, by
\begin{eqnarray}
\label{eq:cpavg}
f_{h^{\prime}}=\frac{f_{h^{\prime}}^{\bar{B}_s}+f_{h^{\prime}}^{B_s}}{2}\,,\quad
A_{CP}^{h^{\prime}}=\frac{f_{h^{\prime}}^{\bar{B}_s}-f_{h^{\prime}}^{B_s}}{f_{h^{\prime}}^{\bar{B}_s}+f_{h^{\prime}}^{B_s}}\,,
\end{eqnarray}
among which only two of the polarization fractions are independent due to the normalization condition $f_{L}+f_{\parallel}+f_{\perp}=1$; such a definition for  $A_{CP}^{h^{\prime}}$ is in fact the same as Eq.~\eqref{eq:cp} for a given $h^{\prime}$. The CP-averaged and CP-violating observables for the relative phases can be constructed, respectively, as
\begin{eqnarray}
\phi_{h^{\prime}}&=&\frac{1}{2}(\phi_{h^{\prime}}^{\bar{B}_s}+\phi_{h^{\prime}}^{B_s})-{\pi}\,{\rm sign}(\phi_{h^{\prime}}^{\bar{B}_s}+\phi_{h^{\prime}}^{B_s})\,\theta(| \phi_{h^{\prime}}^{\bar{B}_s}-\phi_{h^{\prime}}^{B_s}|-\pi)\,,\\
\Delta\phi_{h^{\prime}}&=&\frac{1}{2}(\phi_{h^{\prime}}^{\bar{B}_s}-\phi_{h^{\prime}}^{B_s})+\pi\,\theta(| \phi_{h^{\prime}}^{\bar{B}_s}-\phi_{h^{\prime}}^{B_s}|-\pi)\,,
\end{eqnarray}
for $h^{\prime}=\parallel\,,\perp$. This phase convention for the amplitudes implies $\phi_{h^{\prime}}=\Delta\phi_{h^{\prime}}=0$ at leading order, where all strong phases are zero~\cite{Beneke:2006hg}, while it should be noted that the sign of $A_L$ relative to the transverse amplitudes differs from the experimental convention, which leads to an offset of $\pi$ for $\phi_{\parallel\,,\perp}$~\cite{Beneke:2006hg,Bobeth:2014rra}.

It should be noted that the above ``theoretical" definitions are in the flavor-eigenstate basis and at $t=0$.
The fact complicating the concept of $B_s$ decay observables is caused by the significant effects of time-dependent oscillation between $\bar{B}_{s}$ and $B_{s}$ states. Concerning the decays of $\bar{B}_{s}$ and $B_{s}$  mesons into a common final state, $\bar{f}=f$, the untagged decay rate is the sum of the two time-dependent components, $\Gamma[\bar{B}_s(t)\to f]+\Gamma[B_s(t)\to f]$~\cite{Dunietz:2000cr,DeBruyn:2012wj}, which yields the averaged and time-integrated branching ratio~\cite{DeBruyn:2012wj}
\begin{eqnarray}
\widehat{ {\cal B}}[B_{s}\to f_{h^{\prime}}]=\frac{1}{2}\left( \frac{R_{f_{h^{\prime}}}^H}{\Gamma_s^H}+\frac{R_{f_{h^{\prime}}}^L}{\Gamma_s^L}\right)=\frac{\tau_{B_s}}{2}(R_{f_{h^{\prime}}}^H+R_{f_{h^{\prime}}}^L)\left[\frac{1+y_sH_{f_{h^{\prime}}}}{1-y_s^2}\right]\,,
\end{eqnarray}
where $R_{f_{h^{\prime}}}^{H,L}\equiv \Gamma[B_{s}^{H,L}\to f_{h^{\prime}}]$ with the heavy and light mass-eigenstates,  $|B_{s}^{H,L}\rangle=p|B_{s}\rangle\mp q|\bar{B}_{s}\rangle$;  $\tau_{B_s}\equiv \Gamma_{s}^{-1}=2/(\Gamma_{s}^L+\Gamma_{s}^H)$ is the mean lifetime; $H_{f_{h^{\prime}}}=(R_{f_{h^{\prime}}}^{H}-R_{f_{h^{\prime}}}^{L})/(R_{f_{h^{\prime}}}^{H}+R_{f_{h^{\prime}}}^{L})$ is the CP asymmetry due to the width difference; $y_s$ is the parameter proportional to the width difference
\begin{eqnarray}
y_{s}&\equiv&\frac{\Delta\Gamma_{s}}{2\Gamma_{s}}\equiv
\frac{\Gamma^{L}_{s}-\Gamma^{H}_{s}}{2\Gamma_{s}}\,.
\end{eqnarray}

Then the relation between the experimentally measurable and theoretically calculated branching fractions, $\widehat{ {\cal B}}[B_{s}\to f_{h^{\prime}}]$ and  ${\cal B}[B_{s}\to f_{h^{\prime}}]$, can be written as~\cite{Bobeth:2014rra,DeBruyn:2012wj}
\begin{eqnarray}\label{eq:rela}
\widehat{ {\cal B}}[B_{s}\to f_{h^{\prime}}]=\frac{1+y_s H_{f_{h^{\prime}}}}{1-y_s^2}{\cal B}[B_{s}\to f_{h^{\prime}}]\,,\quad
\widehat{ {\cal B}}[B_{s}\to f]=\sum_{h^{\prime}=L,\parallel\,,\perp}\widehat{ {\cal B}}[B_{s}\to f_{h^{\prime}}]\,.
\end{eqnarray}
Here the decay width parameter $y_s$ is universal for $B_s$ decays and has been well measured, $y_s=0.063\pm0.005$~\cite{Amhis:2014hma}.  However, the CP asymmetry $H_{f_{h^{\prime}}}$ is generally non-universal, not only for various $B_s$ decay modes, but also for various polarization states. Moreover, its values in most of $B_s$ decays are not measured. Therefore, we take the SM prediction~\cite{Bobeth:2014rra}
\begin{eqnarray}
H_{f_{h^{\prime}}}=\frac{2{\rm Re}(\lambda_{f_{h^{\prime}}})}{1+|\lambda_{f_{h^{\prime}}}|^2}\,,\quad
 \lambda_{f_{h^{\prime}}}=\frac{q}{p}\frac{\bar{A}_{f_{h^{\prime}}}}{A_{f_{h^{\prime}}}}\,.
\end{eqnarray}
Accordingly, the experimentally measurable polarization fractions  should also be modified as
\begin{eqnarray}\label{eq:relafh}
\widehat{ f}_{h^{\prime}}=\frac{\widehat{\cal B}[B_{s}\to f_{h^{\prime}}]}{\widehat{ {\cal B}}[B_{s}\to f]}
=\frac{(1+y_s H_{f_{h^{\prime}}}){\cal B}[B_{s}\to f_{h^{\prime}}]}{\sum_{h^{\prime}}(1+y_s H_{f_{h^{\prime}}}) {\cal B}[B_{s}\to f_{h^{\prime}}]}\,;
\end{eqnarray}
they still satisfy the normalization condition  $\widehat{ f}_{L}+\widehat{ f}_{\parallel}+\widehat{ f}_{\perp}=1$. In addition, such a correction induced by the $B_s$ oscillation does not affect the definition for the three polarization-dependent CP asymmetries given by  Eq.~\eqref{eq:cpavg}.

In general, $-1\leqslant H_{f_{h^{\prime}}}\leqslant 1$, and the difference between $\widehat{ {\cal B}}[B_{s}\to f]$ and ${\cal B}[B_{s}\to f]$ can, therefore, reach up to ${\cal O}(10\%)$ for final states that are CP-eigenstates, as has been observed for some cases~\cite{DeBruyn:2012wj}.  On the other hand, for the case of  a flavor-specific decay~($\bar{f}\neq f$), where $H_{f_{h^{\prime}}}=0$, the correction factor in Eq.~\eqref{eq:rela} is simplified as $1/(1-y_s^2)$, which implies a good approximation $\widehat{ {\cal B}}[B_{s}\to f]\simeq {\cal B}[B_{s}\to f]$ due to $y_s^2\sim 4\times10^{-3}\ll 1$. In the following sections, the hat symbol, ``$\widehat{~~}$'', is omitted for convenience.

\section{Numerical results and discussions}

\begin{table}[t]
\begin{center}
\caption{\small Values of the input parameters: Wolfenstein parameters, pole and running quark masses, decay constants, form factors, Gegenbauer moments~and decay width parameter $y_s$.
}
\label{ppvalue}
\vspace*{0.1cm}
\renewcommand{\arraystretch}{1.1}
\tabcolsep 0.18in
\begin{tabular}{l}
\hline\hline
$A=0.8250_{-0.0111}^{+0.0071}$, \quad $\lambda=0.22509_{-0.00028}^{+0.00029}$, \quad$ \bar{\rho}=0.1598_{-0.0072}^{+0.0076}$, \quad $\bar{\eta}=0.3499_{-0.0061}^{+0.0063}$\,; \quad \cite{CKMfitter} \\
\hline
$m_{c}=1.67\pm0.07$~GeV, \quad $m_{b}=4.78\pm0.06$~GeV, \quad  $m_{t}=174.2\pm1.4$ GeV,\\
$\frac{\bar{m}_{s}(\mu)}{\bar{m}_{u,d}(\mu)}=27.3\pm0.7$, \quad  ${\bar{m}}_{s}$(2GeV)$=96_{-4}^{+8}$~MeV, \quad $\bar{m}_{b}(\bar{m}_{b})=4.18_{-0.03}^{+0.04}$~GeV\,; \quad \cite{Olive:2016xmw}\\
\hline
$f_{B_{s}}=227.2\pm3.4$~MeV, \quad $f_{K^{*}}=204\pm7$~MeV, \quad $f_{K^{*}}^{\perp}=159\pm6$~MeV,\\
$f_{\rho}=213\pm5$~MeV, \quad $f_{\rho}^{\perp}=160\pm7$~MeV, \quad  $f_{\phi}=233\pm4$~MeV,\\
$f_{\phi}^{\perp}=191\pm4$~MeV, \quad $f_{\omega}=197\pm8$~MeV, \quad $f_{\omega}^{\perp}=148\pm13$~MeV\,; \quad \cite{Olive:2016xmw,Aoki:2016frl,Straub:2015ica}\\
\hline
$A_{0}^{B_{s}\to\phi}=0.389\pm0.045$, \quad $A_{1}^{B_{s}\to\phi}=0.296\pm0.027$, \quad
$V^{B_{s}\to\phi}=0.387\pm0.033$, \\
$A_{0}^{B_{s}\to K^{*}}=0.314\pm0.048$, \quad
$A_{1}^{B_{s}\to K^{*}}=0.230\pm0.025$, \quad
$V^{B_{s}\to K^{*}}=0.296\pm0.030$\,; \quad \cite{Straub:2015ica}\\
\hline
$a_{1}^{\parallel,\perp}(\phi)=0$,~ $a_{1}^{\parallel,\perp}(\rho)=0$,~
$a_{1}^{\parallel,\perp}(\omega)=0$,~
$a_{1}^{\parallel,(\perp)}(K^*)=0.06\pm0.04(0.04\pm0.03)$, \quad\\
$a_{2}^{\parallel,(\perp)}(\phi)=0.23\pm0.08(0.14\pm0.07)$,\quad
$a_{2}^{\parallel,(\perp)}(\rho)=0.17\pm0.07(0.14\pm0.06)$,\\
$a_{2}^{\parallel,(\perp)}(\omega)=0.15\pm0.12(0.14\pm0.12)$,\quad
$a_{2}^{\parallel,(\perp)}(K^*)=0.16\pm0.09(0.10\pm0.08)$\,;\quad
\cite{Straub:2015ica,Dimou:2012un}\\
\hline
$y_s=\Delta \Gamma_s/(2\Gamma_s)=0.063\pm0.005$\,.~\cite{Amhis:2014hma}\\
\hline\hline
\end{tabular}
\end{center}
\end{table}

\begin{table}[t]
\begin{center}
\caption{\small Experimental data for the measured observables of $\bar B_{s}\to \rho^{0}\phi$, $K^{*0}\bar K^{*0}$, $\phi K^{*0}$ and $\phi \phi$ decays, as well as the deviations of theoretical results from data in cases I and II, {\it i.e.}, the $\chi_i\sigma_i$~($i$ denotes a given observable) evaluated at the best-fit points of $(\rho_{A}^{i,f}, \phi_{A}^{i,f})$ with the other inputs given by Eq.~\eqref{eq:datarhophi} and Table~\ref{ppvalue}. The bold value denotes the largest deviation in the fit. }
\label{pull}
\vspace*{0.2cm}
\renewcommand{\arraystretch}{1.1}
\tabcolsep 0.26in
\begin{tabular}{llcccccc}
\hline\hline
{Observable} &{Decay mode}&{Exp.~\cite{Amhis:2014hma}}&{Case I}&{Case II}\\ \hline
${\cal B}[10^{-6}]$&$\bar B_{s}\to \rho^{0}\phi$&$0.27\pm0.07$&$+0.71\,\sigma$&$+0.71\,\sigma$ \\
&$\bar B_{s}\to K^{*0}\bar K^{*0}$&$10.8\pm2.6$&$+4.31\,\sigma$&$0.00\,\sigma$\\
&$\bar B_{s}\to \phi K^{*0}$&$1.13\pm0.30$&\boldmath{ $-2.57\,\sigma$}&\boldmath{$-3.17\,\sigma$}\\
&$\bar B_{s}\to \phi \phi$&$18.6\pm1.6$&$0.00\,\sigma$&$0.00\,\sigma$\\
\hline
$f_L[\%]$&$\bar B_{s}\to K^{*0}\bar K^{*0}$&$20.1\pm7.0$&$+5.99\,\sigma$&$+0.13\,\sigma$\\
&$\bar B_{s}\to \phi K^{*0}$&$51\pm17$&$+0.41\,\sigma$&$0.00\,\sigma$\\
&$\bar B_{s}\to \phi \phi$&$36.1\pm2.2$&$0.00\,\sigma$&$0.00\,\sigma$\\
\hline
$f_{\bot}[\%]$
&$\bar B_{s}\to K^{*0}\bar K^{*0}$&$58.4\pm8.5$&$-4.96\,\sigma$&$-1.21\,\sigma$\\
&$\bar B_{s}\to \phi K^{*0}$&$28\pm12$&$-0.50\,\sigma$&$0.00\,\sigma$ \\
&$\bar B_{s}\to \phi \phi$&$30.6\pm2.3$&$0.00\,\sigma$&$0.00\,\sigma$\\
\hline
$\phi_{\parallel}+\pi$
&$\bar B_{s}\to \phi K^{*0}$&$1.75^{+0.70}_{-0.61}$&$0.00\,\sigma$&$+0.31\,\sigma$\\
&$\bar B_{s}\to \phi \phi$&$2.59\pm0.15$&$1.27\,\sigma$&$-0.27\,\sigma$\\
\hline\hline
\end{tabular}
\end{center}
\end{table}

With the theoretical formulae given above, we now present our numerical results and discussions. The values of the input parameters used in our evaluation are summarized in Table~\ref{ppvalue}. So far, only some observables of $\bar B_{s}\to$$\rho^0\phi$, $\phi K^{*0}$, $\phi \phi$ and $K^{*0} \bar K^{*0}$ decays, including the CP-averaged branching ratios, the polarization fractions and the relative phases between different helicity amplitudes, have been measured. The experimental data on these measured observables~\cite{Amhis:2014hma} are listed in the ``Exp." column of Table~\ref{pull}, and will be used as constraints in the following $\chi^2$-fits.

In order to probe the HSS and WA contributions in charmless $B_s\to VV$ decays, we perform $\chi^2$-analyses for the end-point parameters, adopting the statistical fitting approach illustrated in our previous work~\cite{Chang:2014rla}~(cf. Appendix C of Ref.~\cite{Chang:2014rla} for detail). In the following fits and posterior predictions, we have to evaluate the theoretical uncertainties induced by the inputs listed in Table~\ref{ppvalue}.  The total theoretical errors are obtained by evaluating separately the uncertainties induced by each input parameter and then adding them in quadrature.

Our $\chi^2$-fits are based on the topology-dependent parametrization scheme for the end-point divergence~\cite{Zhu:2011mm,Wang:2013fya}. This implies that we need four free parameters $(\rho_{A}^{i,f}, \phi_{A}^{i,f})$ (where the superscripts $i$ and $f$, as introduced in section 2, correspond to the nonfactorizable and factorizable annihilation topologies, respectively) to describe the WA contributions. Besides, we also need  two free parameters $(\rho_{H}, \phi_{H})$ to describe the HSS contributions.

\subsection{Constraints on \boldmath{$(\rho_{H}, \phi_{H})$} from \boldmath{$\bar B_{s}\to \rho^0\phi$} decay}

\begin{figure}[t]
\begin{center}
\subfigure[]{
\label{Fig.0.1}
\includegraphics[width=6.2cm]{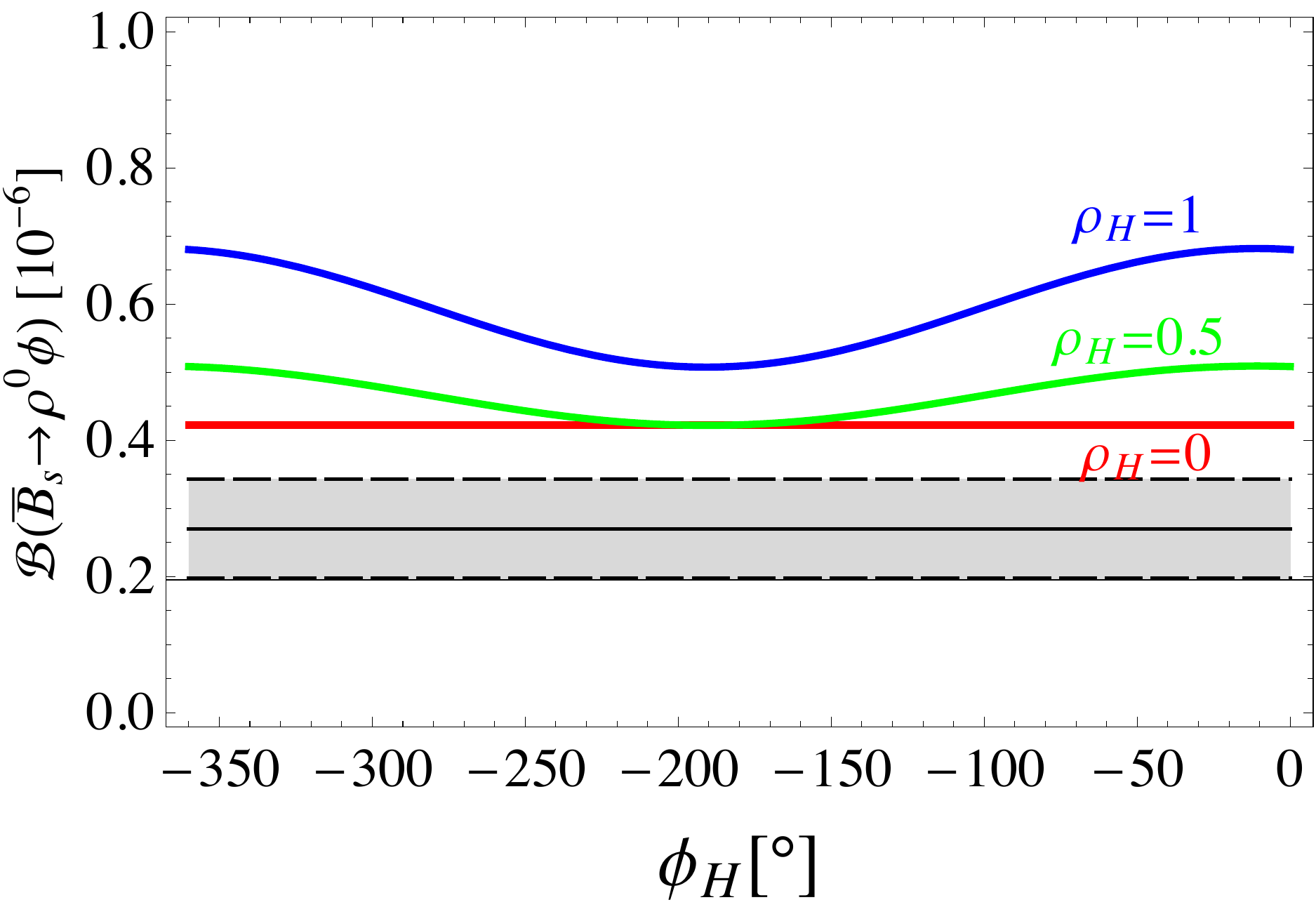}}\qquad
\subfigure[]{
\label{Fig.0.2}
\includegraphics[width=6.6cm]{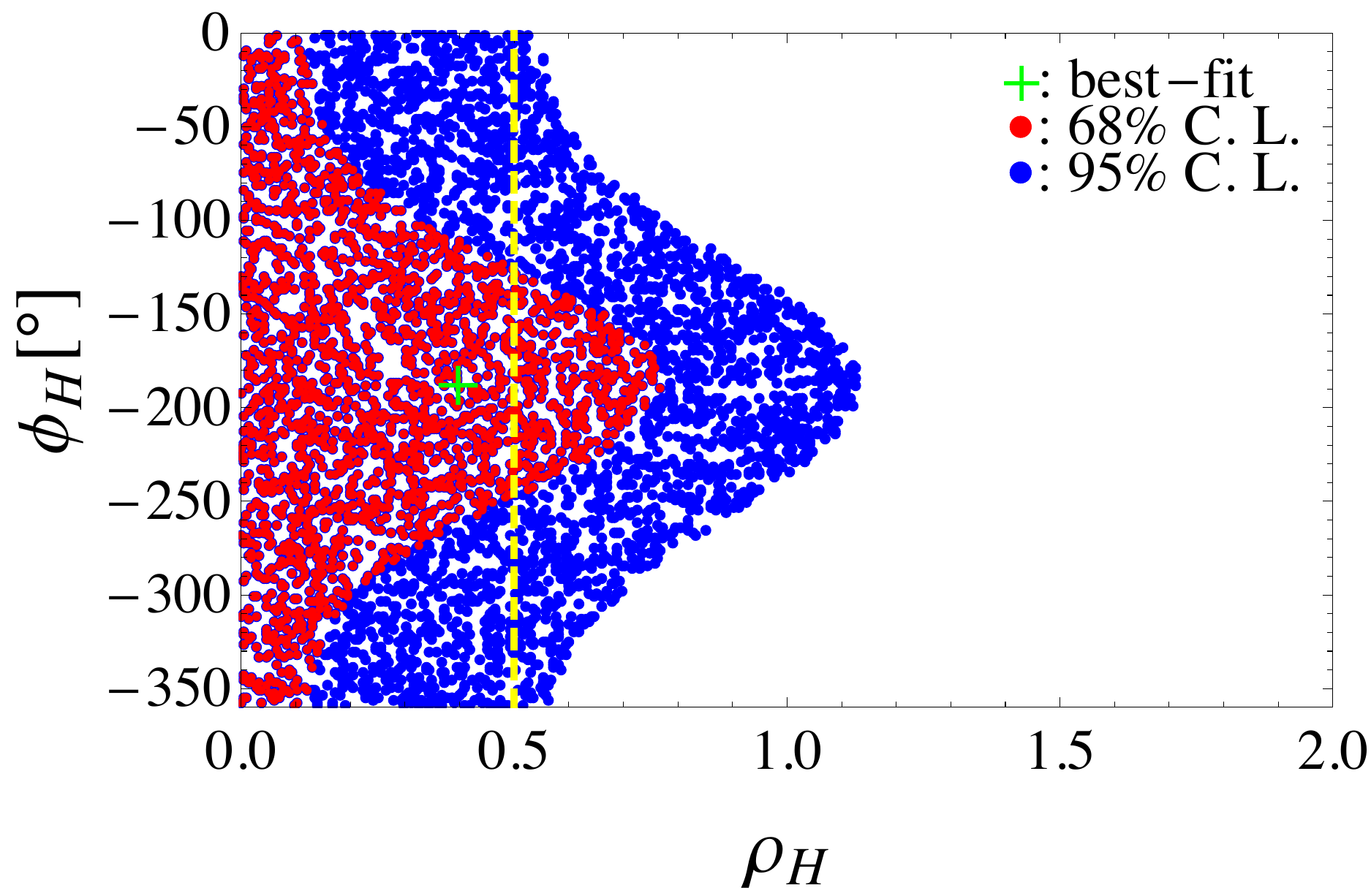}}
\caption{\label{c0} \small Fig.~(a) shows the dependence of ${\cal B}(\bar B_{s}\to \rho^0\phi)$ on $\phi_H$ with different $\rho_H$ labeled in the figure; the gray band is the experimental data within $1\sigma$ error bars.  Fig.~(b) shows the allowed spaces of $(\rho_{H}, \phi_{H})$ at 68\% C.L. and 95\% C.L. under the constraint from the measured ${\cal B}(\bar B_{s}\to \rho^0\phi)$; the dashed line corresponds to $\rho_{H}=0.5$, and the best-fit point corresponds to $\chi^2_{\rm min}= 0.19$. }
\end{center}
\end{figure}

As has been illustrated in Refs.~\cite{Beneke:1999br,Beneke:2000ry,Bobeth:2014rra}, the factor $\rho_H e^{i\phi_H}$ summarizes the remainder of the non-perturbative contribution including a possible strong phase; the numerical size of such a complex parameter is unknown. However, a too large $\rho_H$ will give rise to numerically enhanced subleading $\Lambda_\mathrm{QCD}/m_b$ contributions compared with the formally leading terms. Thus, the size of $\rho_H$ should be carefully coped with.

As analyzed in the last section, the $\bar B_{s}\to \rho^0\phi$ decay is independent of WA contributions and sensitive to HSS corrections. Therefore, it provides an ideal channel for probing the end-point parameters in the HSS amplitudes. Recently, the branching ratio of $\bar B_{s}\to \rho^{0}\phi$ decay has been measured by the LHCb collaboration~\cite{Aaij:2016qnm},
\begin{eqnarray}\label{eq:datarhophi}
{\cal B}(\bar B_{s}\to \rho^0\phi)=(2.7\pm0.7_{\rm stat.} \pm 0.2_{\rm syst.})\times 10^{-7}\,,
\end{eqnarray}
with a significance of about $4\sigma$.

Taking $\rho_H=0\,,0.5\,,1$ and using the central values of input parameters in  Table~\ref{ppvalue}, the dependence of the theoretical result for ${\cal B}(\bar B_{s}\to \rho^0\phi)$ on $\phi_H$ is shown in  Fig.~\ref{Fig.0.1}.  It can be clearly seen that the measured ${\cal B}(\bar B_{s}\to \rho^0\phi)$ presents a very stringent constraint on $\rho_H$; the large $\rho_H$ should obviously be ruled out. The fitted space for $(\rho_H,\phi_H)$ is shown in Fig.~\ref{Fig.0.2}. We find that: (i) The large $\rho_H\gtrsim0.75\,(1.15)$ is excluded at $68\%$~($95\%$) C.L..
(ii) The bound of $\phi_H$ cannot be well determined due to the lack of data for the other observables; however, if $\rho_H\gtrsim0.5$, values of $\phi_H$ around $-180^{\circ}$ are favored.

It has been noted that, besides the $\bar B_{s}\to \rho^0\phi$ decay, the large $\rho_H$ is also disfavored by the color-suppressed tree-dominated  $B_d\to \rho^0 \rho^0$ decay~\cite{Chang:2016ren} even though a large HSS correction with large $\rho_H$ is helpful for explaining the ``$\pi K$ and $\pi\pi$ puzzle"~\cite{Chang:2014rla}. In the following analysis and evaluation, we take a conservative choice that
\begin{eqnarray}\label{eq:datarhophi}
\rho_H=0.5\,,\qquad \phi_H[^{\circ}]=-180\pm100\,,
\end{eqnarray}
as inputs.
Even though such a $\phi_H$ has a large uncertainty, its effect on the following analysis for $(\rho_A^{i,f},\phi_A^{i,f})$ would be not significant because the HSS contribution with $\rho_H=0.5$ is severely suppressed.

\subsection{Case I: constraints on  \boldmath{$(\rho_{A}, \phi_{A})$} from \boldmath{$\bar B_{s}\to \phi K^{*0}$} and \boldmath{$\phi \phi$} decays}

\begin{figure}[t]
\begin{center}
\subfigure[]{
\label{Fig.1.1}
\includegraphics[width=7cm]{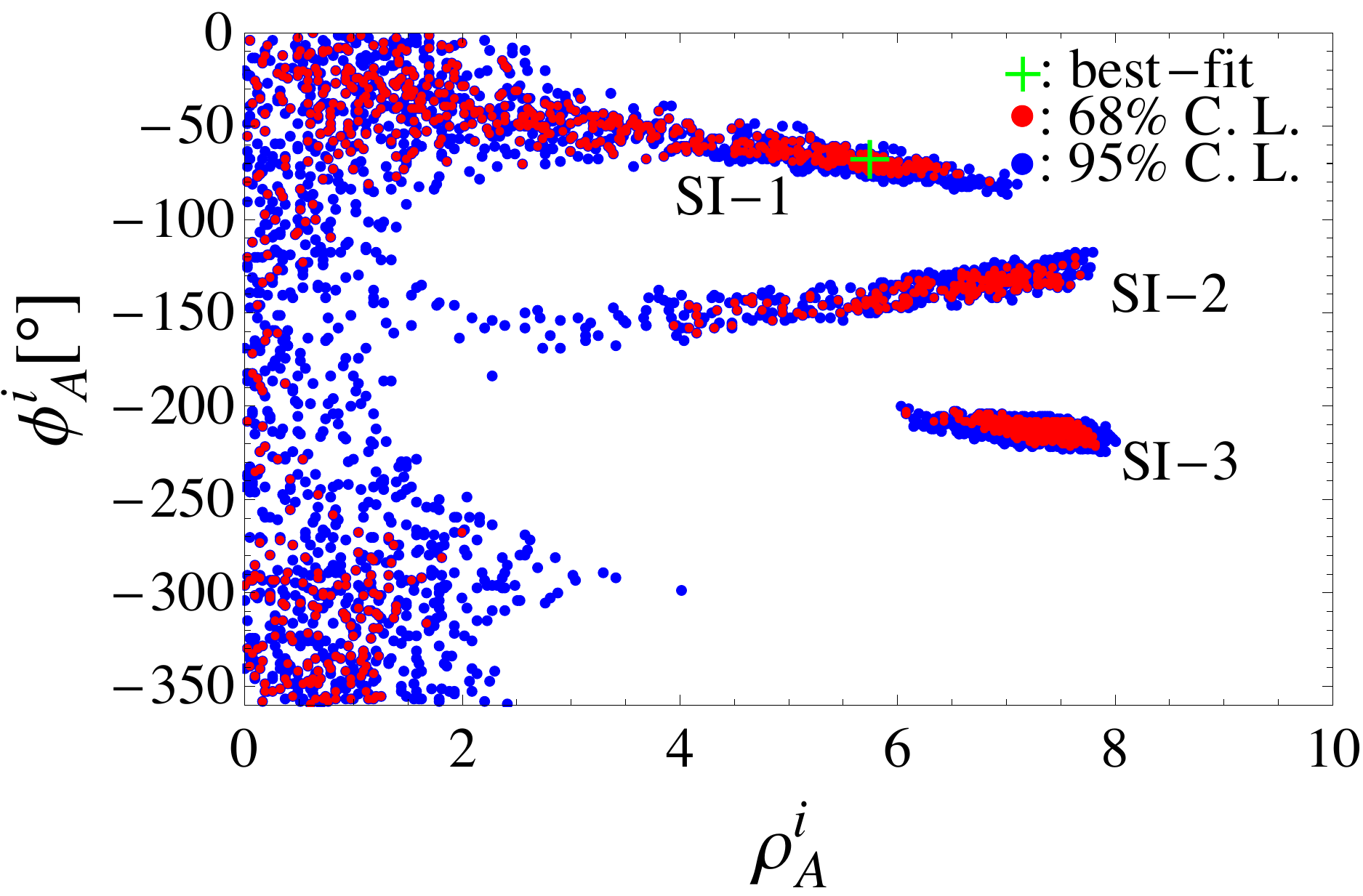}}\qquad
\subfigure[]{
\label{Fig.1.2}
\includegraphics[width=7cm]{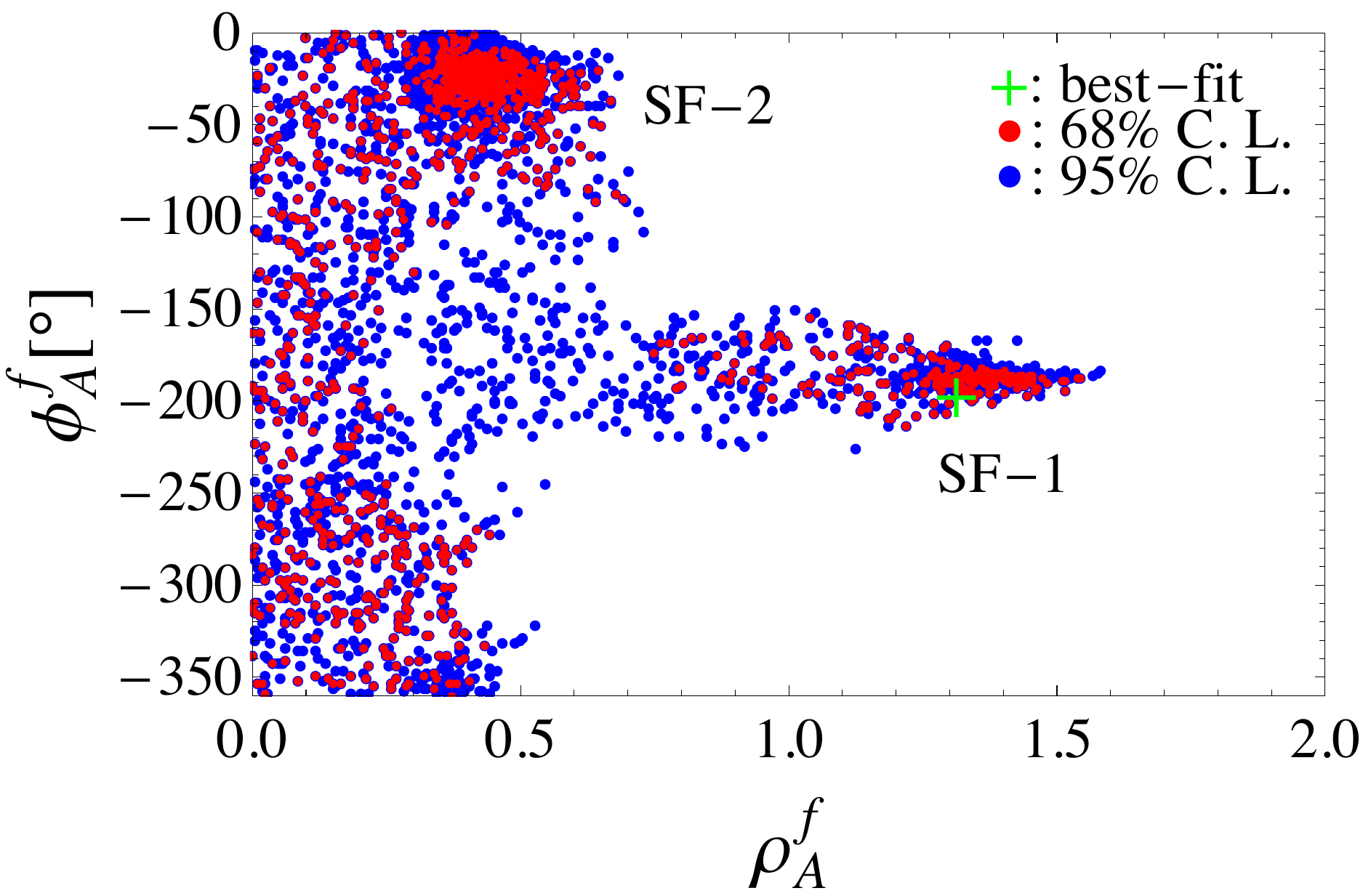}}\qquad\\
\subfigure[]{
\label{Fig.1.3}
\includegraphics[width=7cm]{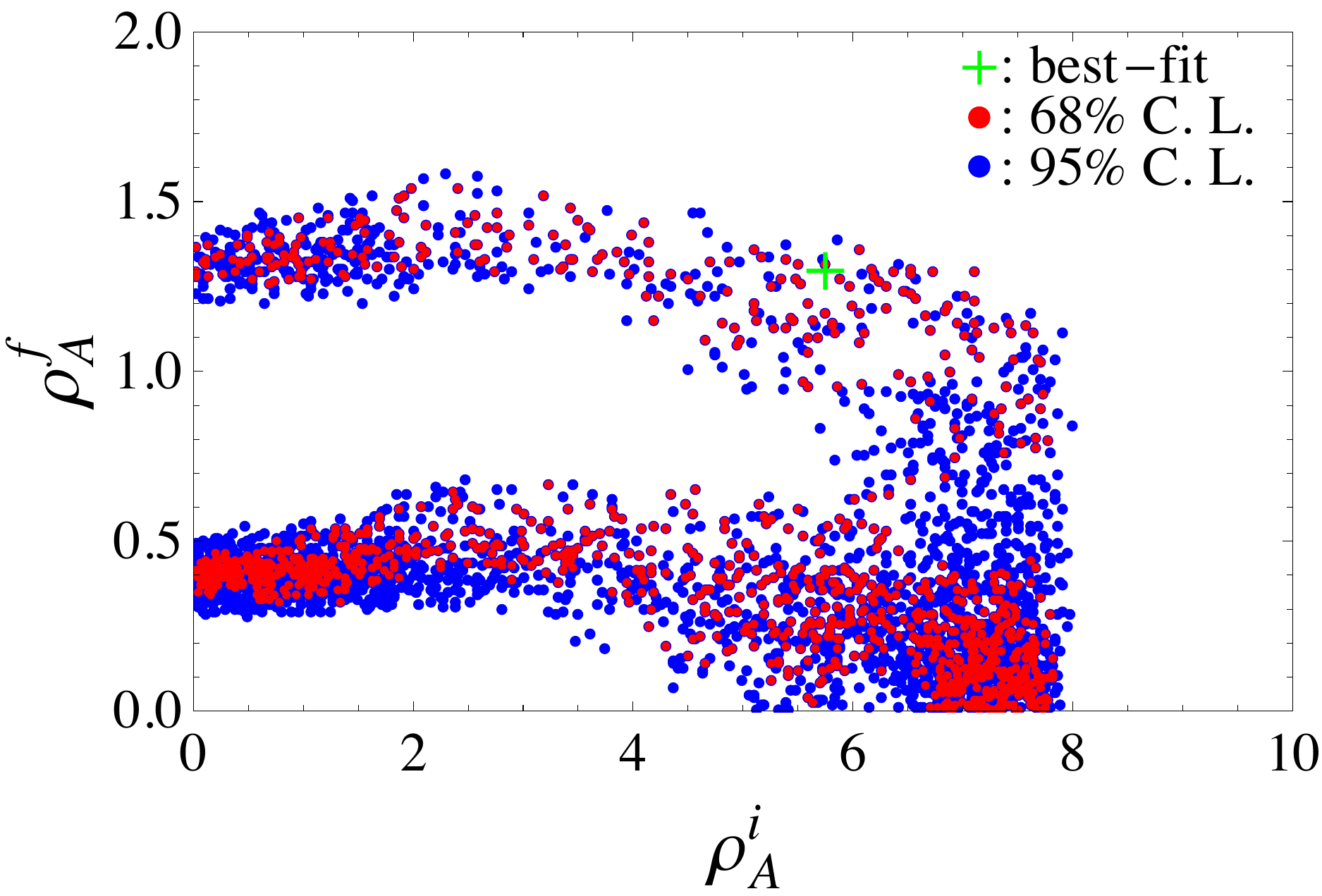}}\qquad
\subfigure[]{
\label{Fig.1.4}
\includegraphics[width=7cm]{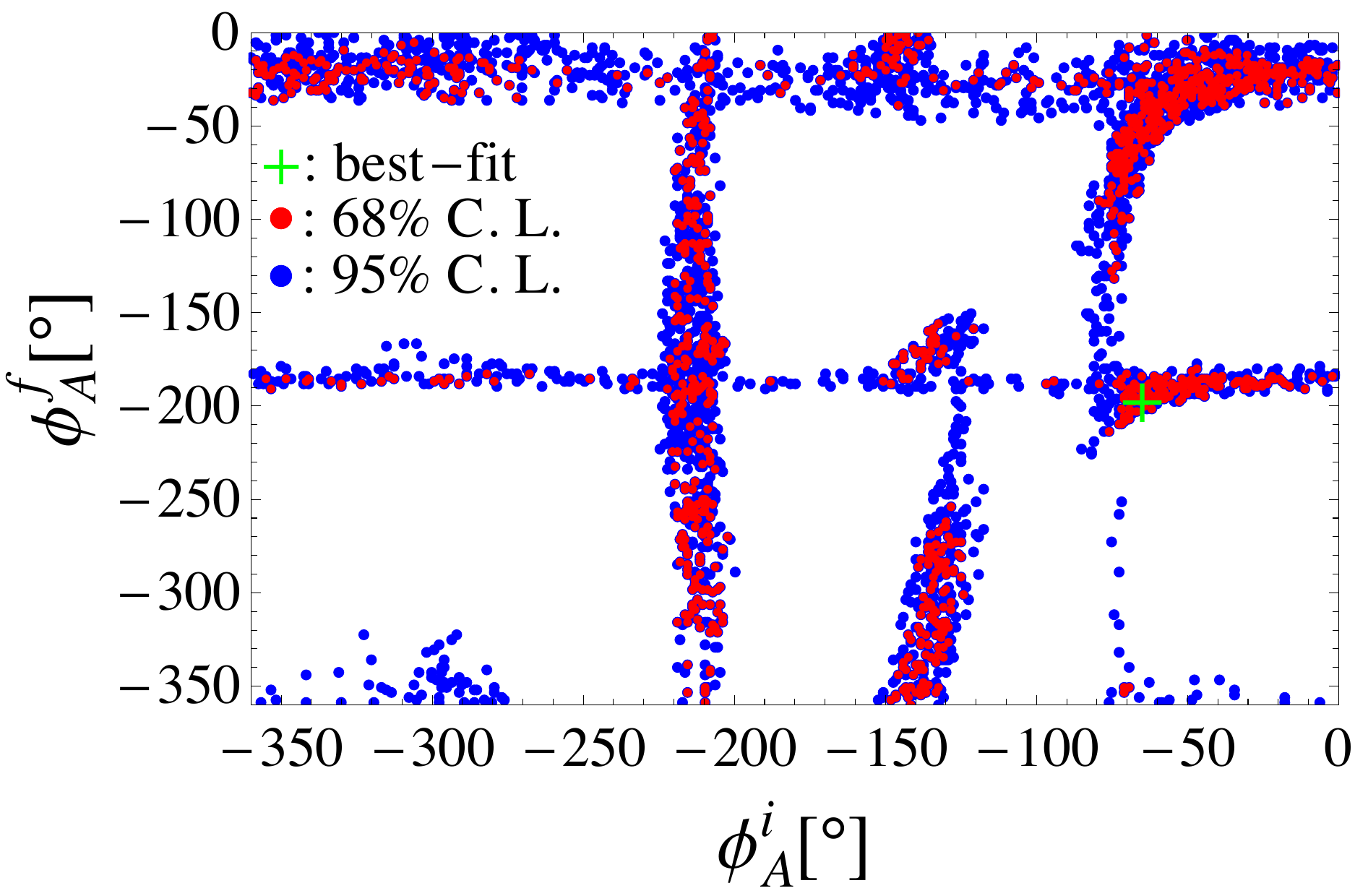}}
\caption{\label{c1} \small The allowed spaces of $(\rho_{A}^{i,f}, \phi_{A}^{i,f})$~( Figs.~(a) and (b) )  at 68\% C.L. and 95\% C.L. under the constraints from the measured $\bar B_{s}\to \phi K^{*0}$ and $\phi \phi$ decays, namely, case I. The fitted spaces are also shown in $(\rho_{A}^i, \rho_{A}^f)$ and $(\phi_{A}^{i}, \phi_{A}^{f})$ planes  (Figs.~(c) and (d) ) in order to show the possible correlation.
The best-fit point corresponds to $\chi^2_{\rm min}/n_{\rm dof}=8.6/4$.
}
\end{center}
\end{figure}

As can be seen from Eqs.~(\ref{eq:Aphik}) and~(\ref{eq:Aphiphi}), the $\bar B_{s}\to \phi K^{*0}$ and $\phi \phi$ decays have similar amplitude structures. However, being a $\Delta D=1$ transition, the $\bar B_{s}\to \phi K^{*0}$ decay amplitude is suppressed by one power of the Wolfenstein parameter $\lambda\simeq0.23$ compared with that of $\bar B_{s}\to \phi \phi$. This explains why the branching ratio ${\cal B}(\bar B_{s}\to \phi K^{*0})$ should be much smaller than ${\cal B}(\bar B_{s}\to \phi \phi)$. One can see from Tables ~\ref{pull} and~\ref{Bs} that the previous QCDF~\cite{Cheng:2009mu,Beneke:2006hg} and pQCD~\cite{Zou:2015iwa} predictions are in good agreement with each other for ${\cal B}(\bar B_{s}\to \phi \phi)$ and ${\cal B}(\bar B_{s}\to \phi K^{*0})$;  in addition, their predictions are consistent with the data for the former, but are much smaller than the data for the latter. Such a deviation could possibly be moderated by the different WA contributions involved in these two decays~(cf. Eqs.~(\ref{eq:Aphik}) and~(\ref{eq:Aphiphi})). To this end, we firstly take the measured $\bar B_{s}\to \phi K^{*0}, \phi \phi$ decays as constraints to fit the WA contributions, which is named as ``case I'' for convenience of discussion.

Under the constraints from the measured $\bar B_{s}\to \phi K^{*0}$ and $\phi \phi$ decays (there are totally eight observables available, see Table~\ref{pull}), the allowed spaces of $(\rho_{A}^{i}, \phi_{A}^{i})$ and $(\rho_{A}^{f}, \phi_{A}^{f})$ are shown in Fig.~\ref{c1}.
We find that:
\begin{itemize}
\item As shown in Figs.~\ref{Fig.1.1} and \ref{Fig.1.2}, the spaces of $(\rho_{A}^{i}, \phi_{A}^{i})$ and $(\rho_{A}^{f}, \phi_{A}^{f})$  are bounded into three and two separate regions, respectively, at 68\% C.L., which are labeled as SI-1, 2, 3 and SF-1, 2 for convenience of discussion. We do not find any direct correspondence between SF-1, 2 for $(\rho_{A}^{f}, \phi_{A}^{f})$ and SI-1, 2, 3 for $(\rho_{A}^{i}, \phi_{A}^{i})$.\footnote{For instance, if we pick out the allowed space of SF-1 at 68\% C.L. for $(\rho_{A}^{f}, \phi_{A}^{f})$, all of the three solutions SI-1,2,3 for $(\rho_{A}^{i}, \phi_{A}^{i})$ are still allowed.}
    
\item For $(\rho_{A}^{f}, \phi_{A}^{f})$, as shown in Fig.~\ref{Fig.1.2}, the space of SF-1 is strictly bounded at 68\% C.L., while the constraint on the one of SF-2 is very loose. The best-fit point with $\chi^2_{\rm min}/n_{\rm dof}=8.6/4$ falls in SF-1; numerically,
    \begin{eqnarray}\label{eq:bff1}
    (\rho_{A}^{f}, \phi_{A}^{f})_{\text{best-fit}}=(1.31,-195^{\circ})\qquad \text{\rm SF-1}\,.
    \end{eqnarray}
    It should be noted that we can also find a point in  SF-2, $(0.45,-40^{\circ})~\text{\rm [SF-2]}$, having a $\chi^2$ value similar to the $\chi^2_{\rm min}$ in  SF-1. The situation for $(\rho_{A}^{i}, \phi_{A}^{i})$ is similar to the one for $(\rho_{A}^{f}, \phi_{A}^{f})$, but is much more complicated as Fig.~\ref{Fig.1.1} shows. The best-fit point in SI-1 is
    \begin{eqnarray}\label{eq:bfi1}
    (\rho_{A}^{i}, \phi_{A}^{i})_{\text{best-fit}}=(5.75,-65^{\circ})~\qquad\text{\rm SI-1}\,.
    \end{eqnarray}
    These allowed spaces in Figs.~\ref{Fig.1.1} and \ref{Fig.1.2} will be  further confronted with the measured observables of $\bar B_{s}\to K^{*0}\bar K^{*0}$ decay in the next subsection.

\item The correlation, $\rho_{A}^{i}\,{\rm vs.}\,\rho_{A}^{f}$, is shown by Fig.~\ref{Fig.1.3}. One can see again that $\rho_{A}^{f}$ is significantly divided into two parts. The relation between $\rho_{A}^{f}$ and $\rho_{A}^{i}$ is not clear due to the large uncertainties except that they can not be equal to zero simultaneously. The correlation, $\phi_{A}^{i}\,{\rm vs.}\,\phi_{A}^{f}$,  shown in Fig.~\ref{Fig.1.4} is very interesting. One can clearly see that the $\phi_{A}^{i}$ can be well determined except when $\phi_{A}^{f}\sim -30^{\circ}$ or $-190^{\circ}$ and vice versa; the case for $\phi_{A}^{f}$ is similar. Hence, the phases $\phi_{A}^{i,f}$ are expected to be well determined when more constraints are considered.
\end{itemize}

As argued in Refs.~\cite{Zhu:2011mm,Wang:2013fya}, the parameters $(\rho_{A}^{f}, \phi_{A}^{f})$ are expected to be universal for $B_{u,d}$ and $B_s$ systems, while  $(\rho_{A}^{i}, \phi_{A}^{i})$ are flavor-dependent on the initial states.  Comparing with the fitted results in  $B_{u,d}$ system~\cite{Chang:2016ren},  we find that the best-fit values, Eqs.~\eqref{eq:bff1} and \eqref{eq:bfi1}, are very similar to the results, $(\rho_{A}^{i}, \phi_{A}^{i})_{B_d}\simeq (5.80, -70^{\circ})$ and $(\rho_{A}^{f}, \phi_{A}^{f})_{B_d}\simeq (1.19, -158^{\circ})$ ( {\it i.e.}, solution C given by Eq.~(14) in Ref.~\cite{Chang:2016ren}) obtained by fitting to $B_{u,d}\to \rho K^{*}$ and $\bar{K}^{*} K^{*}$ decays.
However, it should be noted that the results in Ref.~\cite{Chang:2016ren} are based on the assumption $(\rho_H, \phi_H)=(\rho_{A}^{i}, \phi_{A}^{i})$,
which is not employed in this paper because $\rho_H$ is strictly constrained by $B_s\to \rho\phi$ decay as analyzed in the last subsection; thus, the flavor dependence of $(\rho_{A}^{i}, \phi_{A}^{i})$ is indeterminable here.

The goodness of the fit can be characterized by $\chi^2_{\rm min}/n_{\rm dof}$ and p-value.\footnote{$n_{\rm dof}$ is the number of degrees of freedom given by the number of measurements minus the number of fitted parameters. In the evaluation of p-value, we assume that the goodness-of-fit statistic follows the $\chi^2$ p.d.f~\cite{Olive:2016xmw}.} Numerically, we obtain
\begin{eqnarray}\label{eq:gnf}
\chi^2_{\rm min}/n_{\rm dof}=8.6/4\,,\qquad \text{p-value}=0.07
\end{eqnarray}
at the best-fit point given by Eqs.~\eqref{eq:bff1} and \eqref{eq:bfi1}.  In order to find which observables lead to the large $\chi^2_{\rm min}$ and small p-value, we summarize the deviations of theoretical results from data for the considered observables in the fourth column of Table~\ref{pull}. It can be clearly seen that the tension between the  theoretical result  and data for ${\cal B}(\bar B_{s}\to \phi K^{*0})$, $\sim -2.57\,\sigma$, dominates the contributions to $\chi^2_{\rm min}$. Numerically, one can find $\chi^{2}_{{\cal B}(\bar B_{s}\to \phi K^{*0})}/ \chi^2_{\rm min}= 77\%$. Such a tension implies that the problem of large ${\cal B}(\bar B_{s}\to \phi K^{*0})$ mentioned in the beginning of this subsection is hardly to be moderated by the WA contribution due to the constraints from the other measured observables. In addition, from Table~\ref{pull}, we also find some significant tensions
in the $\bar B_{s}\to K^{*0}\bar K^{*0}$ decay, which is not considered in the fit in this case~(case I). This implies that the best-fit points given by Eqs.~\eqref{eq:bff1} and \eqref{eq:bfi1} might be excluded when the constraints from $\bar B_{s}\to K^{*0}\bar K^{*0}$ decay are considered; and the other fitted spaces in this case  may also suffer challenges from $\bar B_{s}\to K^{*0}\bar K^{*0}$ decay, which will be studied in detail in the next two subsections.

\subsection{Polarizations in \boldmath{$\bar B_{s}\to K^{*0}\bar K^{*0}$} decay \label{sec:3.3}}

\begin{figure}[t]
\begin{center}
\subfigure[]{
\label{Fig.3.1}
\includegraphics[width=6cm]{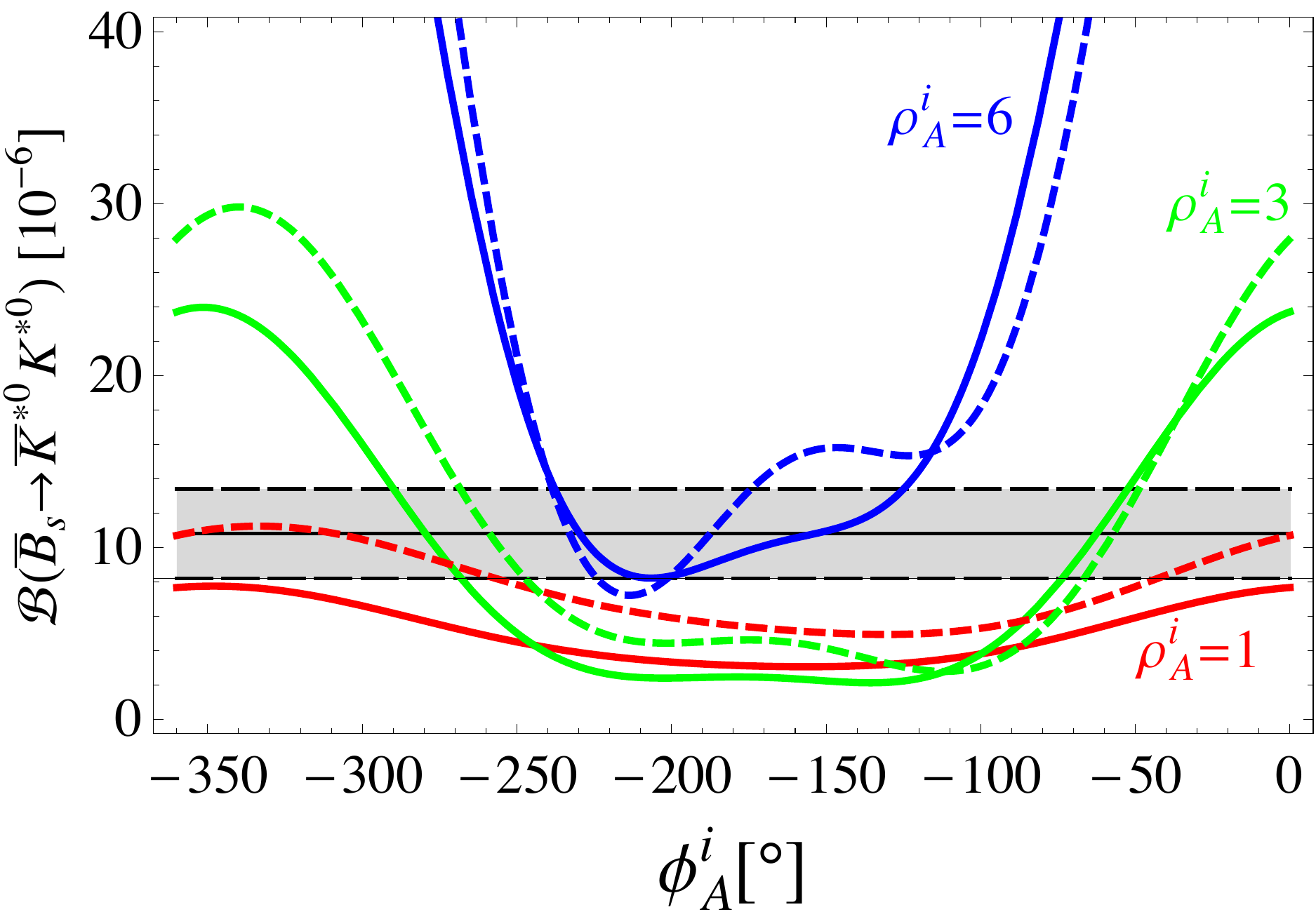}}\quad
\subfigure[]{
\label{Fig.3.2}
\includegraphics[width=6cm]{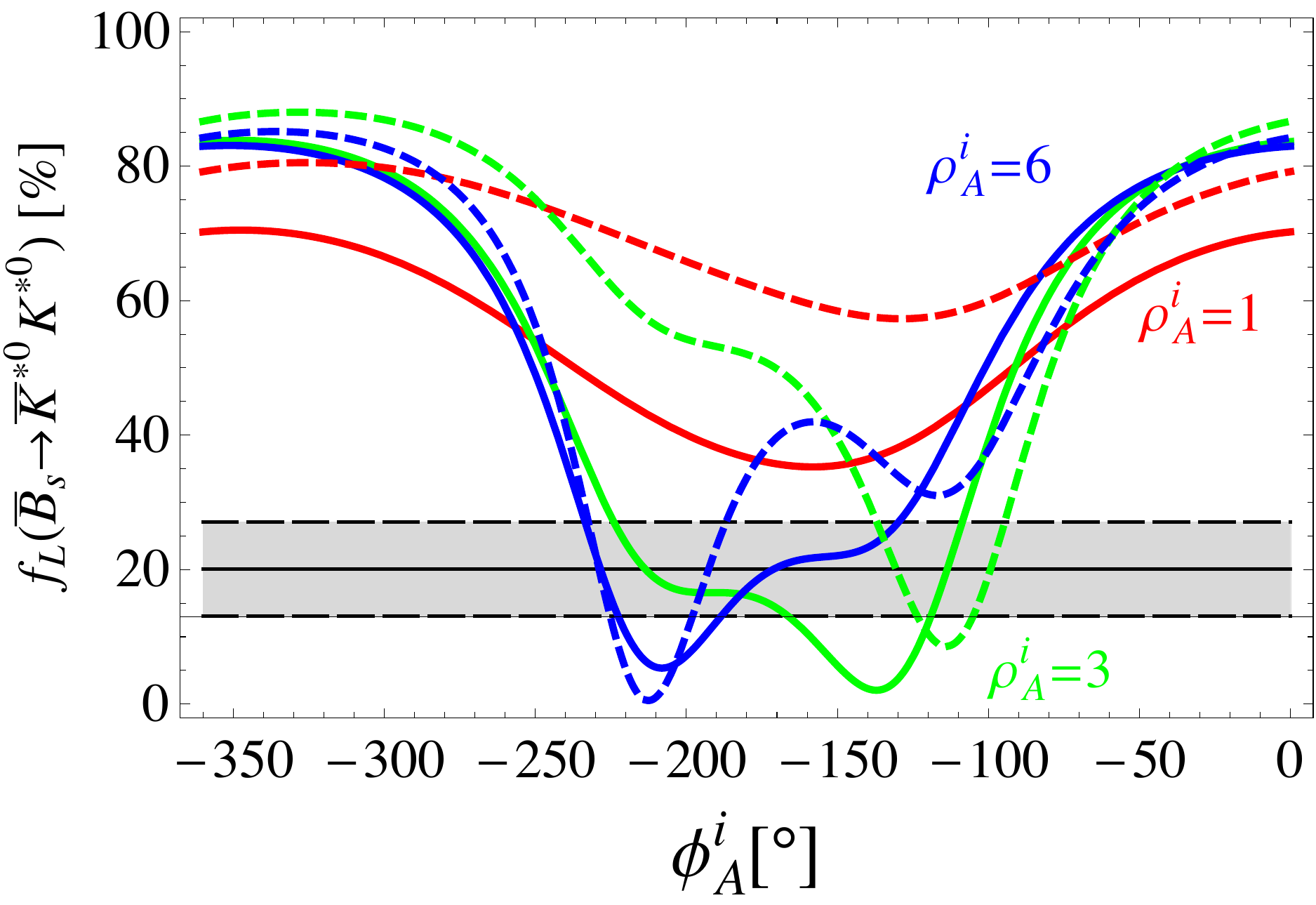}}\quad
\subfigure[]{
\label{Fig.3.3}
\includegraphics[width=6cm]{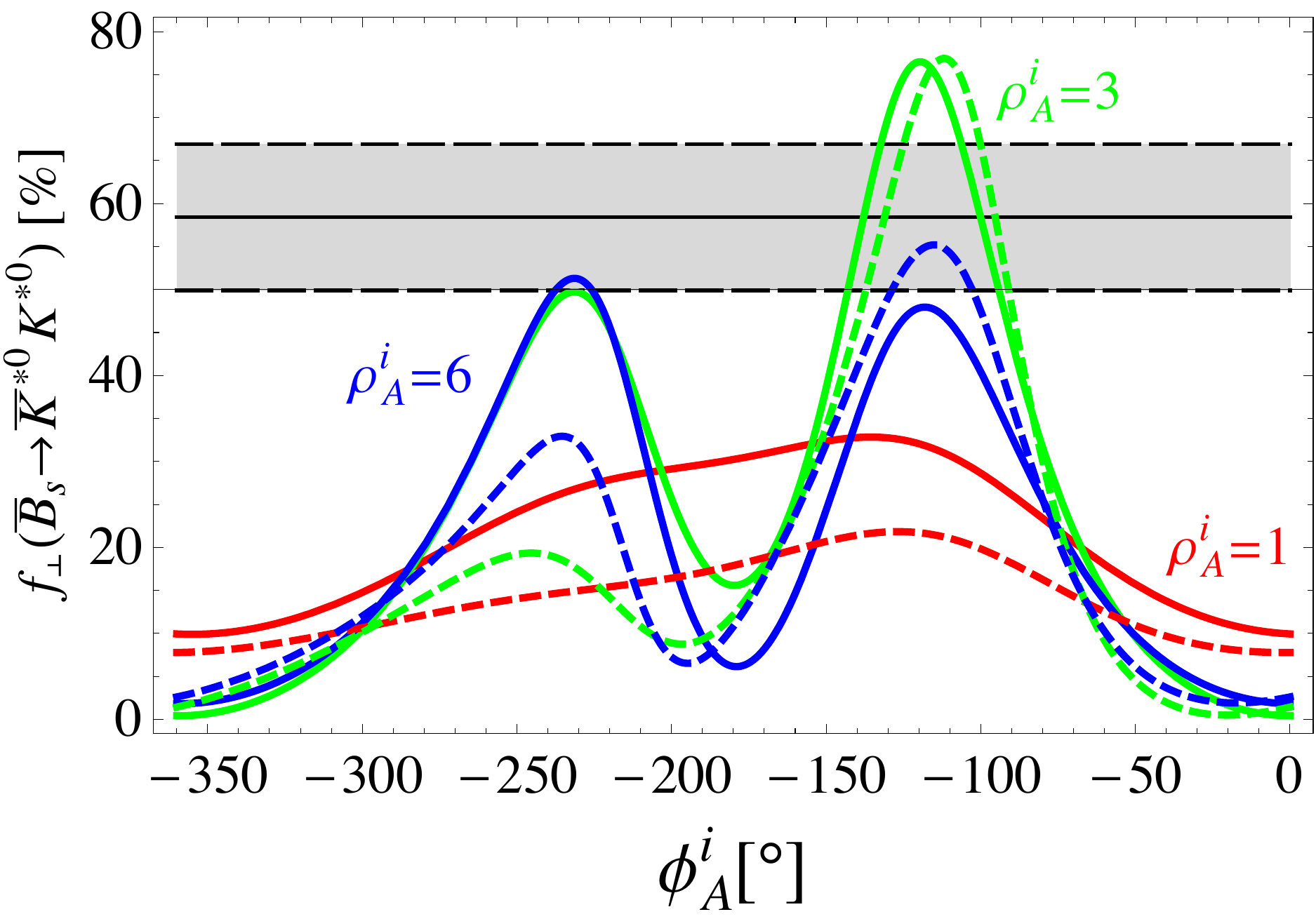}}
\caption{\label{kkdep1} \small
The dependences of ${\cal B}(\bar B_{s}\to K^{*0}\bar K^{*0})$ and $f_{L,\bot}(\bar B_{s}\to K^{*0}\bar K^{*0})$ on the parameters $(\rho_{A}^{i}, \phi_{A}^{i})$, with fixed  $(\rho_{A}^{f}, \phi_{A}^{f})=(1.31,-195^\circ)$~(solid lines) and $(0.45,-40^\circ)$~(dashed lines). The shaded bands are the experimental data within $1~\sigma$ error bars.
}
\end{center}
\end{figure}

In the fit of case I, we do not include the measured $\bar B_{s}\to K^{*0}\bar K^{*0}$ decay, because it is difficult to understand its polarizations measured  by the LHCb collaboration~\cite{Aaij:2015kba}. It is well-known that, due to the $(V-A)$ nature of the SM weak interactions, the hierarchical pattern among the three helicity amplitudes, $A_0:A_{-}:A_+=1:\frac{\Lambda_{\rm QCD}}{m_b}:(\frac{\Lambda_{\rm QCD}}{m_b})^2$, is expected in charmless $B\to VV$ decays~\cite{Kagan:2004uw}. Even after the QCD corrections are taken into account, the charmless $B\to VV$ decay amplitudes are generally still dominated by the longitudinal polarization component. For the penguin-dominated $B\to VV$ decays, the longitudinal polarization fraction is generally predicted at the level of about $50\%$, for instance in the $B\to\phi K^*$ decays~\cite{Aubert:2004xc,Ladisa:2004bp,Hou:2004vj,Li:2004mp,Kim:2004wq,Chang:2006dh,Li:2003hea,Li:2004ti}. Consistent with the above expectation, the longitudinal~(transverse) polarization fraction, $f_{L(\bot)}(\bar B_{s}\to K^{*0}\bar K^{*0})\sim 50\%~(25\%)$, is predicted both in the QCDF~\cite{Beneke:2006hg} and in the pQCD approach~\cite{Zou:2015iwa}.
However, the obviously different experimental results have been measured by the LHCb collaboration~\cite{Aaij:2015kba},
\begin{eqnarray}
f_{L}(\bar B_{s}\to K^{*0}\bar K^{*0})&=&(20.1\pm5.7\pm 4.0)\%\,,\\
f_{\parallel}(\bar B_{s}\to K^{*0}\bar K^{*0})&=&(21.5\pm4.6 \pm1.5)\%\,,
\end{eqnarray}
which imply $f_{\bot}(\bar B_{s}\to K^{*0}\bar K^{*0})=(58.4 \pm8.5)\%$.
Furthermore, these measurements are also inconsistent with the previous theoretical expectation, $f_{\parallel}\approx f_{\bot}$ (the relation $|f_{\parallel}- f_{\bot}|\lesssim 4\%$ is satisfied by most of the charmless $B\to VV$ decays~\cite{Beneke:2006hg}). As a consequence, these possible anomalies  present a challenge to the current theoretical predictions.  Therefore, we would like to check if the modifications of end-point parameters could reconcile these anomalies.

In Fig.~\ref{kkdep1}, we plot the dependences of ${\cal B}(\bar B_{s}\to K^{*0}\bar K^{*0})$ and
$f_{L,\bot}(\bar B_{s}\to K^{*0}\bar K^{*0})$ on the parameters $(\rho_{A}^{i}, \phi_{A}^{i})$ with  $(\rho_{A}^{f}, \phi_{A}^{f})=(1.31,-195^\circ)$ and $(0.45,-40^\circ)$, which are the best-fit values in SF-1 and -2,  respectively, in case I.
From Fig.~\ref{kkdep1}, one can see that: 
\begin{itemize}
\item The small $\rho_{A}^{i}\lesssim 1$ is perhaps allowed by the measured ${\cal B}(\bar B_{s}\to K^{*0}\bar K^{*0})$ with $\phi_{A}^{i}$ around $0^{\circ}$~(or $-360^{\circ}$) as Fig.~\ref{Fig.3.1} shows, but is excluded by both $f_{L}(\bar B_{s}\to K^{*0}\bar K^{*0})$ and $f_{\bot}(\bar B_{s}\to K^{*0}\bar K^{*0})$ as Figs.~\ref{Fig.3.2} and \ref{Fig.3.3} show.
\item With $\rho_{A}^{i}\sim 3$, ${\cal B}(\bar B_{s}\to K^{*0}\bar K^{*0})$ and  $f_{L,\bot}(\bar B_{s}\to K^{*0}\bar K^{*0})$ present different requirements for $\phi_{A}^{i}$, which can be seen by comparing Fig.~\ref{Fig.3.1} with Figs.~\ref{Fig.3.2} and \ref{Fig.3.3}. This implies that the choice  $\rho_{A}^{i}\sim 3$ is also excluded by the anomalies of  $f_{L,\bot}(\bar B_{s}\to K^{*0}\bar K^{*0})$.
\item Only a large $\rho_A^i\sim 6$ with the phase $\phi_{A}^{i}\sim -240^{\circ}$ or $-130^{\circ}$ could possibly account for the current LHCb measurements for $f_{L,\bot}(\bar B_{s}\to K^{*0}\bar K^{*0})$. It is very interesting that such possible solutions are similar to SI-2 and -3 shown in Fig.~\ref{Fig.1.1}. However, the SI-1 is possibly excluded by $f_{L,\bot}(\bar B_{s}\to K^{*0}\bar K^{*0})$. In order to further check such possible solutions, in the next subsection we will perform a combined fit for the end-point parameters with the measured observables of $\bar B_{s}\to \phi K^{*0}$, $\phi \phi$ and $K^{*0}\bar K^{*0}$ decays as constraints.
\end{itemize}

\subsection{Case II: combined fit to the measured \boldmath{$B_s\to VV$} decays}

\begin{figure}[t]
\begin{center}
\subfigure[]{
\label{Fig.4.1}
\includegraphics[width=7cm]{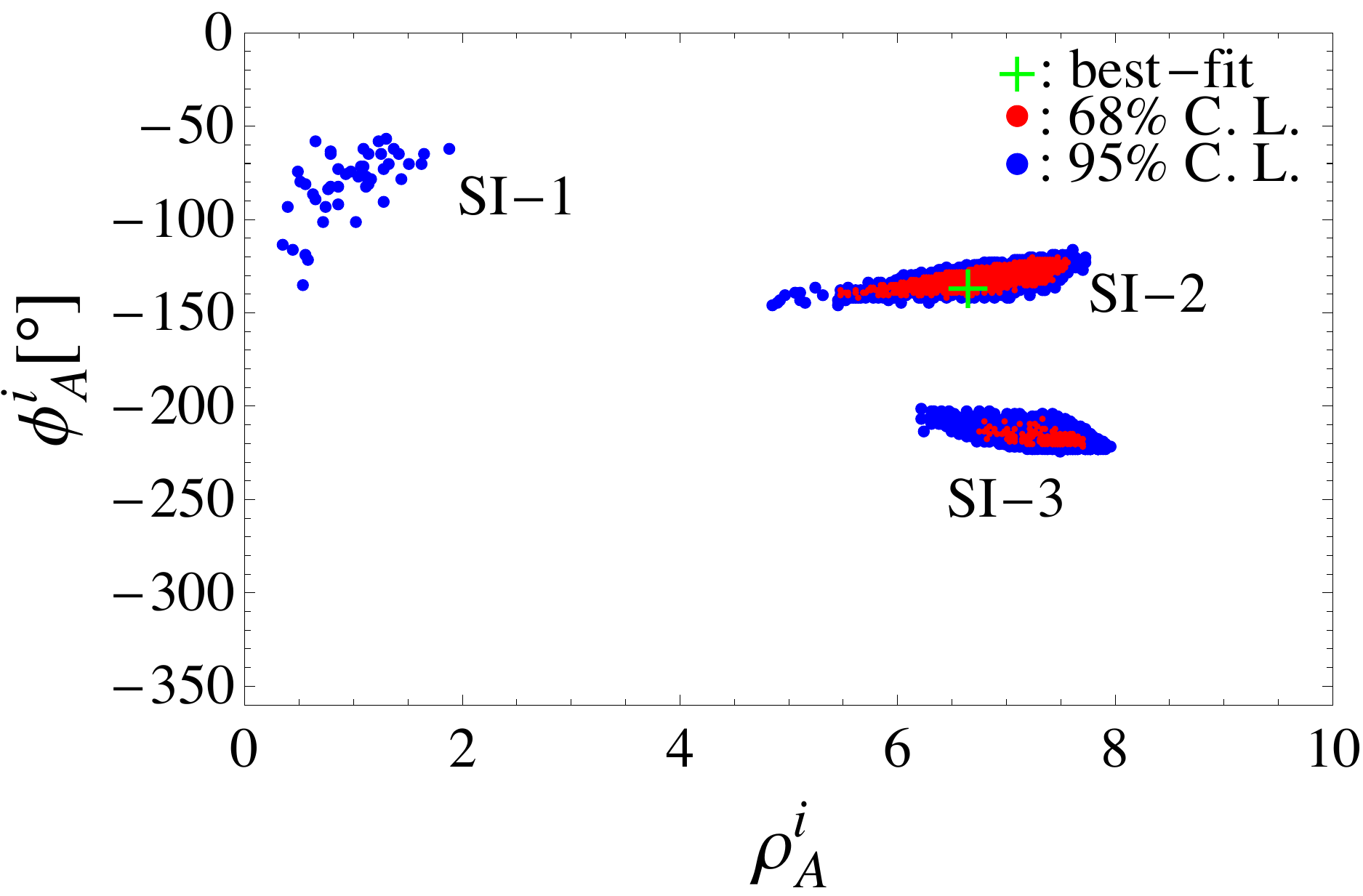}}\quad
\subfigure[]{
\label{Fig.4.2}
\includegraphics[width=7.1cm]{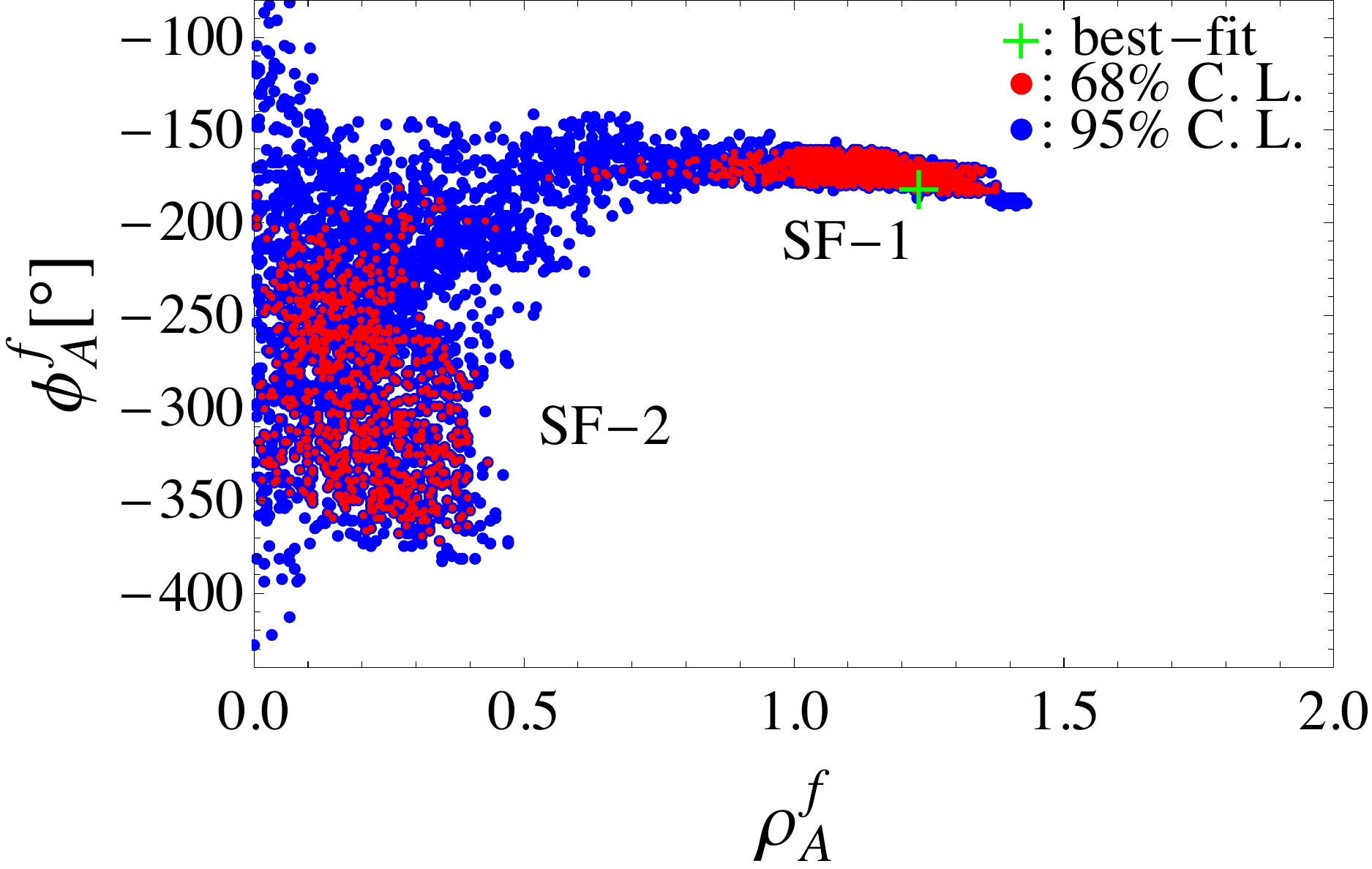}}\\
\subfigure[]{
\label{Fig.4.3}
\includegraphics[width=7cm]{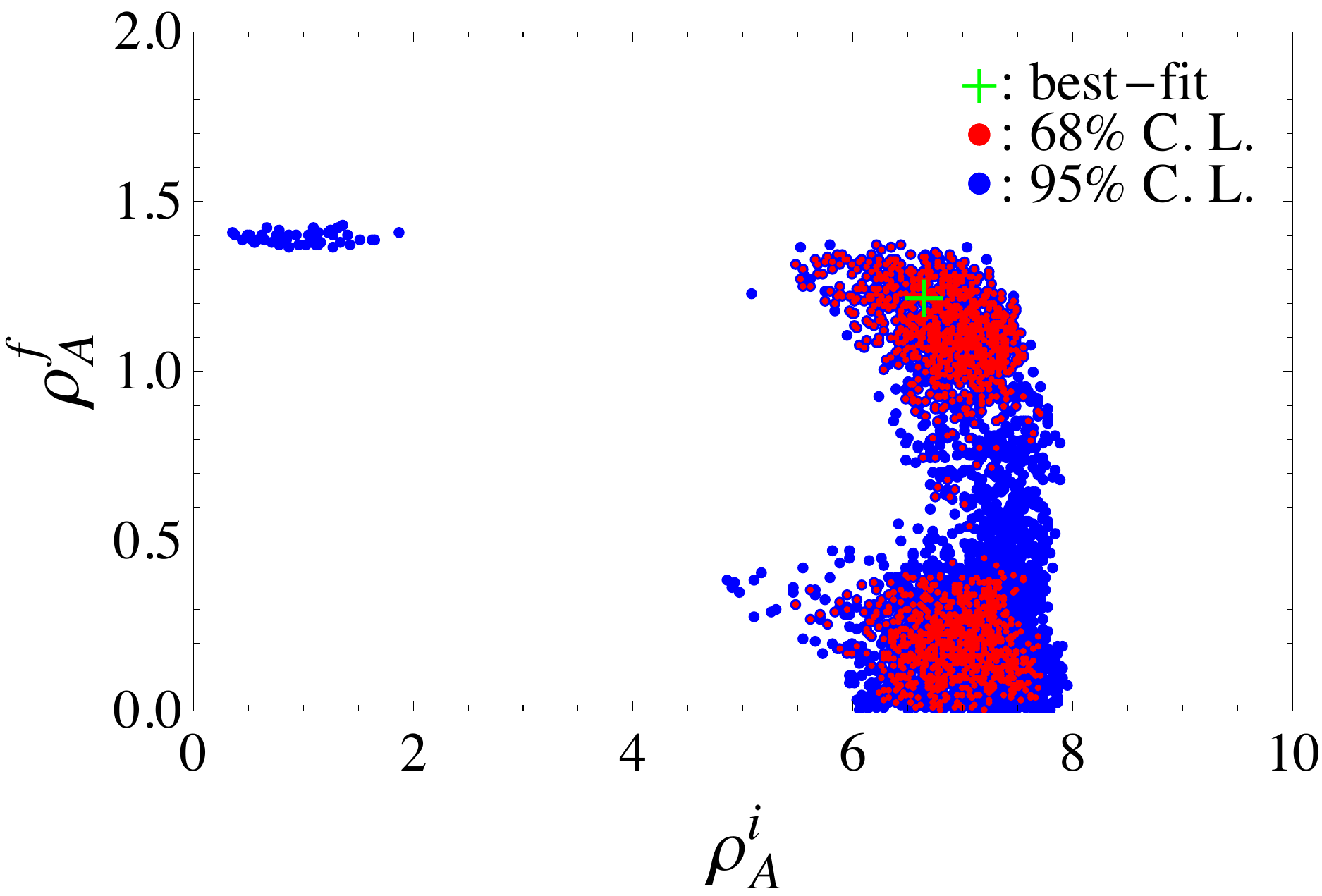}}\quad
\subfigure[]{
\label{Fig.4.4}
\includegraphics[width=7.2cm]{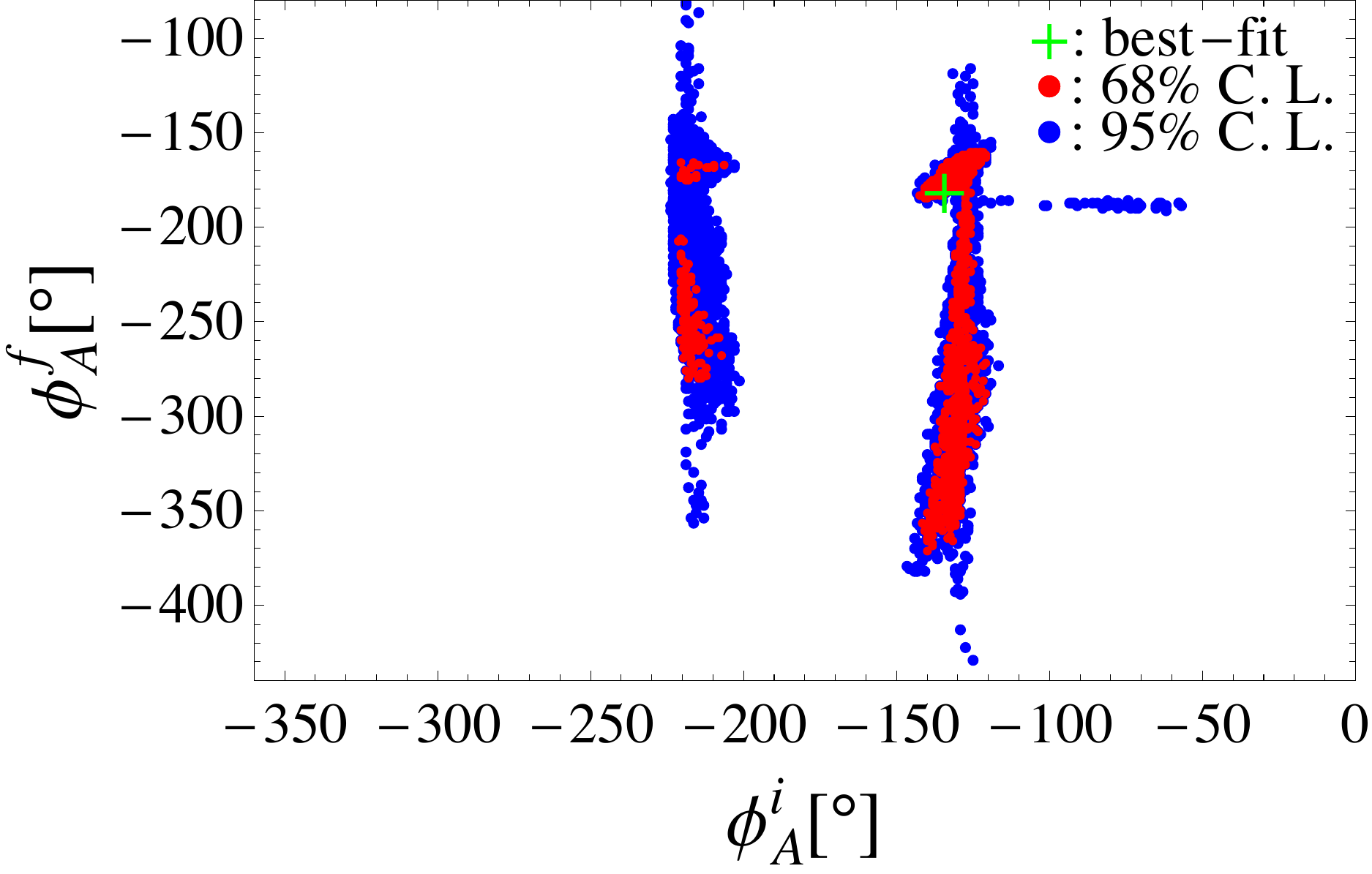}}
\caption{\label{c1pp} \small The allowed spaces of $(\rho_{A}^{i,f}, \phi_{A}^{i,f})$ at $68\%$ C.L. and $95\%$ C.L. under the combined constraints from $\bar B_{s}\to \phi K^{*0}$, $\phi \phi$ and $K^{*0}\bar K^{*0}$ decays, namely, case II. The best-fit points correspond to $\chi_{min}^{2}/n_{\rm dof}=11.2/7$.}
\end{center}
\end{figure}

Under the combined constraints from $\bar B_{s}\to \phi K^{*0}$, $\phi \phi$ and  $K^{*0}\bar K^{*0}$ decays (\textit{i.e.}, the measured 11 observables of $B_s\to VV$ decays are now included), our fitted results for the end-point parameters are shown in Fig.~\ref{c1pp}, which is named as ``case II'' for convenience of discussion. It can be seen that:
\begin{itemize}
\item For $(\rho_{A}^{i}, \phi_{A}^{i})$,  because of the constraints from $f_{L,\bot}(\bar B_{s}\to K^{*0}\bar K^{*0})$, the space SI-1 favored in case I is excluded at 68\% C.L. in case II, which can be clearly seen from Fig.~\ref{Fig.4.1} and easily understood  from the analysis in the subsection~\ref{sec:3.3}, while both  SI-2 and -3 survive, but they are further restricted; the large $\rho_{A}^{i}$ is required to fit $f_{L,\bot}(\bar B_{s}\to K^{*0}\bar K^{*0})$.  The spaces SI-2 and -3 are located symmetrically at two sides of  $\phi_{A}^{i}\sim-180^{\circ}$, and thus lead to a similar $|X_{A}^i|$ but different signs of ${\rm Im}[X_{A}^i]$. In view of $\chi^2_{\rm min}$, SI-2 involving the best-fit point is much more favored. Numerically, we obtain
    \begin{equation} \label{scasIIi}
    (\rho_A^{i},\phi_A^{i}{[^\circ]})=(6.65^{+0.91}_{-1.17}\,,-134^{+13}_{-8})\,.\qquad \text{SI-2}
    \end{equation}
    
\item For $(\rho_{A}^{f}, \phi_{A}^{f})$, the two spaces, SF-1 and -2, are still allowed and further restricted in case II as Fig.~\ref{Fig.4.2} shows.  Such two spaces at 68\% C.L. can be clearly distinguished according to if $\rho_{A}^{f}\gtrsim0.5$. The best-fit point falls in the space SF-1. For SF-1, the ranges of both  $\rho_{A}^{f}$ and  $\phi_{A}^{f}$ are strictly bounded; numerically, we obtain
    \begin{equation} \label{scasIIf}
    (\rho_A^{f},\phi_A^{f}{[^\circ]})=(1.23^{+0.15}_{-0.62}\,,-179^{+19}_{-5})\,.\qquad \text{SF-1}
    \end{equation}
    For  SF-2, even though the space of $(\rho_{A}^{f}, \phi_{A}^{f})$ is further restricted compared with case I, the constraints are still very loose.
    
\item The correlations,  $\rho_{A}^{i}\,{\rm vs.}\,\rho_{A}^{f}$ and $\phi_{A}^{i}\,{\rm vs.}\,\phi_{A}^{f}$, are shown by Figs.~\ref{Fig.4.3} and \ref{Fig.4.4}, respectively. The main difference between case I and case II is that the ranges $\rho_{A}^{i}\lesssim5.5$ and  $\phi_{A}^{i}\not\approx -140^{\circ}$ or $-220^{\circ}$ are excluded  by the $\bar B_{s}\to K^{*0}\bar K^{*0}$ decay at 68\% C.L. in  case II. \\
    In addition, it also can be found that the relation $(\rho_{A}^{i}, \phi_{A}^{i})\neq(\rho_{A}^{f}, \phi_{A}^{f})$ is required at 68\% C.L., which implies that the end-point parameters are topology-dependent and, therefore, confirms the suggestion proposed in Refs.~\cite{Zhu:2011mm,Wang:2013fya}.
\end{itemize}

In this case, we obtain
\begin{eqnarray}\label{eq:gnf2}
\chi^2_{\rm min}/n_{\rm dof}=11.2/7\,,\qquad \text{p-value}=0.13
\end{eqnarray}
at the best-fit point. The deviations of the theoretical results from data are summarized  in the fifth column of Table~\ref{pull}. We again find that the observable ${\cal B}(\bar B_{s}\to \phi K^{*0})$ results in the large $\chi^2_{\rm min}$ and small p-value. Numerically, one can find $\chi^{2}_{{\cal B}(\bar B_{s}\to \phi K^{*0})}/ \chi^2_{\rm min}= 90\%$, which is similar to case I. If we disregard ${\cal B}(\bar B_{s}\to \phi K^{*0})$, we can find
\begin{eqnarray}\label{eq:gnf2}
\chi^2_{\rm min}/n_{\rm dof}=1.65/6\,,\qquad \text{p-value}=0.95\,.
\end{eqnarray}

Comparing case II with case I, we find from Table~\ref{pull} that their main difference is the deviations for the observables of  $\bar B_{s}\to K^{*0}\bar K^{*0}$ decay:  $0.00\,\sigma$~vs.~$+4.31\,\sigma$ for ${\cal B}$, $+0.13\,\sigma$~vs.~$+5.99\,\sigma$ for $f_L$, and $-1.21\,\sigma$~vs.~$-4.96\,\sigma$ for $f_{\bot}$, which again indicates that the abnormal data for $\bar B_{s}\to K^{*0}\bar K^{*0}$ decay can be explained through the WA contributions. However,  in both cases I and II, the deviations for ${\cal B}(\bar B_{s}\to \phi K^{*0})$, $-2.57\,\sigma$ and $-3.17\,\sigma$, are very large; this implies that the measured large ${\cal B}(\bar B_{s}\to \phi K^{*0})=(1.13\pm0.30)\times 10^{-6}$~\cite{Amhis:2014hma}, which is much larger than all of the current predictions, $\sim 0.4 \times 10^{-6}$, in pQCD~\cite{Zou:2015iwa} and QCDF~\cite{Beneke:2006hg,Cheng:2009mu},  is hardly to be accommodated by the WA contributions due to the constraints from the other observables and decay modes.

\subsection{Updated results for charmless \boldmath{$B_s\to VV$} decays}

\begin{table}[htbp]
\begin{center}
\caption{\small Theoretical results for the measured $\bar B_{s}\to K^{*0}\bar K^{*0}$, $\phi K^{*0}$ and $\phi \phi$ decays. The first uncertainty is caused by the input parameters listed in Table~\ref{ppvalue} and $\phi_{H}$ given by Eq.~\eqref{eq:datarhophi}, and the second one arises from the uncertainties of $(\rho_{A}^{i,f}, \phi_{A}^{i,f})$ given by Eqs.~\eqref{scasIIi} and \eqref{scasIIf}.}
\label{Bs}
\vspace*{0.2cm}
\renewcommand{\arraystretch}{1.0}
\footnotesize \tabcolsep 0.08in
\begin{tabular}{llcccccc}
\hline\hline
{Observable} &{Decay mode}&{Case I}&{Case II}&pQCD~\cite{Zou:2015iwa}  &QCDF~\cite{Beneke:2006hg}&QCDF~\cite{Cheng:2009mu}\\ \hline
${\cal B}[10^{-6}]$
&$\bar B_{s}\to K^{*0}\bar K^{*0}$
&$23.3^{+1.3}_{-1.3}$&$10.2^{+1.6+8.1}_{-0.9-6.3}$&$5.4^{+3.0}_{-2.4}$ &$9.1^{+0.5+11.3}_{-0.4-6.8}$&$6.6^{+1.1+1.9}_{-1.4-1.7}$\\
&$\bar B_{s}\to \phi K^{*0}$
&$0.22^{+0.14}_{-0.11}$&$0.11^{+0.07+0.06}_{-0.04-0.01}$&$0.39^{+0.20}_{-0.17}$&$0.4^{+0.1+0.5}_{-0.1-0.3}$&$0.37^{+0.06+0.24}_{-0.05-0.20}$\\
&$\bar B_{s}\to \phi \phi$
&$19.2^{+3.3}_{-1.5}$&$18.4^{+7.8+11.9}_{-1.4-10.6}$&$16.7^{+8.9}_{-7.1}$ &$21.8^{+1.1+30.4}_{-1.1-17.0}$&$16.7^{+2.6+11.3}_{-2.1-8.8}$\\
\hline
$A_{CP}[\%]$
&$\bar B_{s}\to K^{*0}\bar K^{*0}$
&$-0.1^{+0.0}_{-0.0}$&$0.1^{+0.1+0.2}_{-0.1-0.3}$&$0.0$ &$1^{+0+1}_{-0-0}$&$0.4^{+0.8+0.6}_{-0.5-0.4}$\\
&$\bar B_{s}\to \phi K^{*0}$
&$-29.0^{+9.0}_{-3.0}$&$-36.1^{+7.1+21.3}_{-5.2-4.2}$&$0.0$&$-17^{+4+9}_{-5-9}$&$-9^{+3+4}_{-1-6}$ \\
&$\bar B_{s}\to \phi \phi$
&$0^{+0}_{-0}$&$0.6^{+0.2+0.4}_{-0.2-0.4}$&$0.0$ &$1^{+0+1}_{-0-0}$&$0.2^{+0.4+0.5}_{-0.3-0.2}$\\
\hline
$A_{CP}^0[\%]$
&$\bar B_{s}\to K^{*0}\bar K^{*0}$
&$0^{+0}_{-0}$&$-0.2^{+0.2+0.7}_{-0.2-0.8}$&$0.0$ &$0^{+0+0}_{-0-0}$&---\\
&$\bar B_{s}\to \phi K^{*0}$
&$-2.6^{+1.5}_{-2.2}$&$-21.1^{+12.2+26.2}_{-26.9-35.5}$&$0.0$&$-9^{+2+16}_{-3-20}$&--- \\
&$\bar B_{s}\to \phi \phi$
&$-0.9^{+0.2}_{-0.7}$&$0.3^{+0.2+2.0}_{-0.3-0.6}$&$0.0$ &$0^{+0+1}_{-0-0}$&---\\
\hline
$A_{CP}^{\bot}[\%]$
&$\bar B_{s}\to K^{*0}\bar K^{*0}$
&$0.4^{+0.1}_{-0.1}$&$0.2^{+0.1+0.3}_{-0.1-0.3}$&$0.0$ &---&---\\
&$\bar B_{s}\to \phi K^{*0}$
&$8.7^{+3.3}_{-3.7}$&$19.5^{+5.1+1.4}_{-10.2-8.1}$&$0.0$&--- &---\\
&$\bar B_{s}\to \phi \phi$
&$0.6^{+0.1}_{-0.2}$&$0^{+0.2+0.7}_{-0.2-0.4}$&$0.0$ &---&---\\
\hline
$f_L[\%]$
&$\bar B_{s}\to K^{*0}\bar K^{*0}$
&$67.1^{+1.2}_{-5.1}$&$27.7^{+8.2+9.5}_{-6.7-18.9}$&$38.3^{+12.1}_{-10.5}$ &$63^{+0+42}_{-0-29}$&$56^{+4+22}_{-7-26}$\\
&$\bar B_{s}\to \phi K^{*0}$
&$72.0^{+5.1}_{-14.0}$&$43.6^{+14.6+51.5}_{-24.0-25.3}$&$50.0^{+8.1}_{-7.2}$&$40^{+1+67}_{-1-35}$&$43^{+2+21}_{-2-18}$ \\
&$\bar B_{s}\to \phi \phi$
&$35.0^{+4.3}_{-19.0}$&$39.4^{+7.3+14.4}_{-6.6-29.1}$&$34.7^{+8.9}_{-7.1}$ &$43^{+0+61}_{-0-34}$&$36^{+3+23}_{-4-18}$\\
\hline
$f_{\bot}[\%]$
&$\bar B_{s}\to K^{*0}\bar K^{*0}$
&$15.2^{+1.1}_{-1.0}$&$41.8^{+6.3+10.2}_{-5.7-14.0}$&$30.0^{+5.3}_{-6.1}$ &---&---\\
&$\bar B_{s}\to \phi K^{*0}$
&$13.4^{+8.6}_{-2.1}$&$25.9^{+8.4+14.4}_{-9.1-23.5}$&$24.2^{+3.6}_{-3.9}$&---&--- \\
&$\bar B_{s}\to \phi \phi$
&$30.0^{+4.2}_{-2.6}$&$33.0^{+7.0+17.2}_{-4.1-21.3}$&$31.6^{+3.5}_{-4.4}$ &---&---\\
\hline
$\phi_{\parallel}+\pi$
&$\bar B_{s}\to K^{*0}\bar K^{*0}$
&$2.81^{+0.10}_{-0.11}$&$2.20^{+0.19+0.64}_{-0.19-0.40}$&$2.12^{+0.21}_{-0.25}$ &$2.84^{+0+1.00}_{-0-0.54}$&---\\
&$\bar B_{s}\to \phi K^{*0}$
&$1.82^{+0.46}_{-0.52}$&$2.77^{+0.66+1.33}_{-0.80-0.56}$&$1.95^{+0.21}_{-0.22}$&$2.71^{+0+1.00}_{-0-0.54}$&--- \\
&$\bar B_{s}\to \phi \phi$
&$3.09^{+0.26}_{-0.31}$&$2.38^{+0.17+1.13}_{-0.44-0.44}$&$2.01\pm0.23$ &$2.80^{+0+0.75}_{-0-0.50}$&---\\
\hline
$\phi_{\perp}+\pi$
&$\bar B_{s}\to K^{*0}\bar K^{*0}$
&$2.61^{+0.14}_{-0.15}$&$2.17^{+0.22+0.22}_{-0.20-0.32}$ &---&---\\
&$\bar B_{s}\to \phi K^{*0}$
&$1.10^{+0.32}_{-0.32}$&$3.51^{+0.92+1.40}_{-1.10-0.48}$&--- &---\\
&$\bar B_{s}\to \phi \phi$
&$4.70^{+0.27}_{-0.31}$&$0.53^{+0.24+0.61}_{-0.24-0.35}$ &---&---\\
\hline
$\Delta\phi_{\parallel}$
&$\bar B_{s}\to K^{*0}\bar K^{*0}$
&$0^{+0}_{-0}$&$0.01^{+0+0.01}_{-0-0}$ &$0^{+0+0.09}_{-0-0.03}$&---\\
&$\bar B_{s}\to \phi K^{*0}$
&$-0.07^{+0.05}_{-0.05}$&$0.17^{+0.34+0.16}_{-0.19-0.20}$&$0.03^{+0+0.15}_{-0-0.06}$&--- \\
&$\bar B_{s}\to \phi \phi$
&$-0.01^{+0}_{-0}$&$0^{+0+0.01}_{-0-0.01}$ &$0^{+0+0}_{-0-0}$&---\\
\hline
$\Delta\phi_{\perp}$
&$\bar B_{s}\to K^{*0}\bar K^{*0}$
&$0^{+0}_{-0}$&$0.01^{+0+0.01}_{-0-0}$ &---&---\\
&$\bar B_{s}\to \phi K^{*0}$
&$-0.16^{+0.05}_{-0.05}$&$0.26^{+0.19+0.09}_{-0.24-0.06}$&---&--- \\
&$\bar B_{s}\to \phi \phi$
&$-0.01^{+0}_{-0}$&$0.01^{+0+0.01}_{-0-0.01}$ &---&---\\
\hline\hline
\end{tabular}
\end{center}
\end{table}

\begin{table}[htbp]
\begin{center}
\caption{\small Theoretical results for the $b\to d$ induced $\bar B_{s}\to \rho^{-}K^{*+}$, $\rho^{0}K^{*0}$ and $\omega K^{*0}$ decays. The other captions are the same as in Table~\ref{Bs}.} \label{deld}
\vspace*{0.2cm}
\renewcommand{\arraystretch}{1.0}
\footnotesize \tabcolsep 0.05in
\begin{tabular}{llccccccc}
\hline\hline
{Observable} &{Decay mode}&{Case I}&{Case II}&pQCD~\cite{Zou:2015iwa}  &QCDF~\cite{Beneke:2006hg}&QCDF~\cite{Cheng:2009mu}\\ \hline
${\cal B}[10^{-6}]$
&$\bar B_{s}\to \rho^{-}K^{*+}$
&$28.5^{+9.8}_{-8.3}$&$28.5^{+9.8}_{-8.4}$$^{+0.1}_{-0.1}$
&$24.0^{+10.9+1.2+0.0}_{-8.7-1.4-2.4}$ &$25.2^{+1.5+4.7}_{-1.7-3.1}$&$21.6^{+1.3+0.9}_{-2.8-1.5}$\\
&$\bar B_{s}\to \rho^{0}K^{*0}$
&$1.23^{+1.20}_{-0.63}$&$1.20^{+1.20}_{-0.57}$$^{+0.01}_{-0.01}$
&$0.40^{+0.19+0.11+0.00}_{-0.15-0.07-0.03}$&$1.5^{+1.0+3.1}_{-0.5-1.5}$&$1.3^{+2.0+1.7}_{-0.6-0.3}$ \\
&$\bar B_{s}\to \omega K^{*0}$
&$1.16^{+1.07}_{-0.51}$&$1.14^{+1.05}_{-0.51}$$^{+0.03}_{-0.04}$
&$0.35^{+0.16+0.09+0.04}_{-0.14-0.08-0.08}$ &$1.2^{+0.7+2.3}_{-0.3-1.1}$&$1.1^{+1.5+1.3}_{-0.5-0.3}$\\
\hline
$A_{CP}[\%]$
&$\bar B_{s}\to \rho^{-}K^{*+}$
&$-9.1^{+1.0}_{-1.1}$&$-20.1^{+0.5}_{-0.6}$$^{+3.3}_{-1.3}$
&$-9.1^{+1.4+1.0+0.2}_{-1.5-1.2-0.3}$ &$-3^{+1+2}_{-1-3}$&$-11^{+1+4}_{-1-1}$\\
&$\bar B_{s}\to \rho^{0}K^{*0}$
&$-36.6^{+14.5}_{-17.0}$&$-18.0^{+11.3}_{-12.8}$$^{+3.7}_{-9.6}$
&$62.7^{+6.4+10.5+7.5}_{-5.9-16.0-7.9}$&$27^{+5+34}_{-7-27}$&$46^{+15+10}_{-17-25}$\\
&$\bar B_{s}\to \omega K^{*0}$
&$-36.5^{+12.3}_{-16.6}$&$-18.6^{+8.6}_{-14.4}$$^{+9.1}_{-3.6}$
&$-78.1^{+2.9+13.1+8.1}_{-2.2-7.4-8.3}$ &$-34^{+10+31}_{-7-43}$&$-50^{+20+21}_{-15-6}$\\
\hline
$A_{CP}^0[\%]$
&$\bar B_{s}\to \rho^{-}K^{*+}$
&$-1.5^{+0.5}_{-0.8}$&$-0.2^{+0.1}_{-0.1}$$^{+1.2}_{-0.4}$
&$-2.71^{+0.68}_{-0.72}$ &$-2^{+1+6}_{-0-3}$&---\\
&$\bar B_{s}\to \rho^{0}K^{*0}$
&$3.8^{+8.6}_{-6.6}$&$-5.4^{+13.3}_{-13.1}$$^{+4.5}_{-5.2}$
&$-17.5^{+21.2}_{-13.0}$&$-5^{+1+49}_{-0-18}$&---\\
&$\bar B_{s}\to \omega K^{*0}$
&$-5.3^{+5.0}_{-6.9}$&$3.1^{+8.8}_{-9.0}$$^{+4.3}_{-4.5}$
&$-5.99^{+23.52}_{-50.21}$ &$6^{+1+19}_{-1-60}$&---\\
\hline
$A_{CP}^{\bot}[\%]$
&$\bar B_{s}\to \rho^{-}K^{*+}$
&$35.4^{+3.8}_{-5.5}$&$-6.1^{+1.1}_{-1.5}$$^{+7.8}_{-6.3}$
&$55.0^{+10.3}_{-10.5}$ &---&---\\
&$\bar B_{s}\to \rho^{0}K^{*0}$
&$-34.5^{+41.8}_{-62.4}$&$46.3^{+52.0}_{-53.1}$$^{+26.4}_{-26.6}$
&$22.0^{+29.9}_{-31.4}$&--- &---\\
&$\bar B_{s}\to \omega K^{*0}$
&$44.4^{+52.6}_{-17.6}$&$-33.1^{+58.1}_{-57.0}$$^{+32.3}_{-25.3}$
&$6.95^{+27.91}_{-32.14}$ &---&---\\
\hline
$f_L[\%]$
&$\bar B_{s}\to \rho^{-}K^{*+}$
&$94.7^{+1.6}_{-3.1}$&$94.8^{+1.6}_{-2.9}$$^{+0.2}_{-0.1}$
&$95^{+1+1+0}_{-1-1-0}$ &$92^{+1+5}_{-1-8}$&$92^{+1+1}_{-2-3}$\\
&$\bar B_{s}\to \rho^{0}K^{*0}$
&$83.5^{+10.2}_{-11.7}$&$85.0^{+9.5}_{-10.6}$$^{+1.2}_{-0.8}$
&$57^{+6+6+1}_{-10-8-0}$&$93^{+2+5}_{-3-54}$ &$90^{+4+3}_{-5-23}$\\
&$\bar B_{s}\to \omega K^{*0}$
&$84.5^{+10.4}_{-11.5}$&$86.4^{+9.6}_{-10.1}$$^{+0.6}_{-0.6}$
&$50^{+7+11+1}_{-8-15-1}$ &$93^{+2+5}_{-4-49}$&$90^{+3+3}_{-4-23}$\\
\hline
$f_{\bot}[\%]$
&$\bar B_{s}\to \rho^{-}K^{*+}$
&$2.4^{+1.4}_{-0.8}$&$2.3^{+1.4}_{-0.8}$$^{+0.1}_{-0.1}$
&$2.31^{+0.22}_{-0.21}$ &---&---\\
&$\bar B_{s}\to \rho^{0}K^{*0}$
&$8.5^{+6.2}_{-5.4}$&$7.1^{+5.0}_{-4.5}$$^{+0.5}_{-0.6}$
&$22.5^{+7.3}_{-4.7}$&---&--- \\
&$\bar B_{s}\to \omega K^{*0}$
&$8.1^{+6.2}_{-5.6}$&$6.4^{+4.9}_{-4.7}$$^{+0.4}_{-0.2}$
&$26.1^{+9.8}_{-7.0}$ &---&---\\
\hline
$\phi_{\parallel}+\pi$
&$\bar B_{s}\to \rho^{-}K^{*+}$
&$3.17^{+0.20}_{-0.22}$&$3.18^{+0.20}_{-0.22}$$^{+0.21}_{-0.23}$
&$3.07^{+0.07}_{-0.09}$ &$3.13^{+0.02+0.18}_{-0.02-0.18}$&---\\
&$\bar B_{s}\to \rho^{0}K^{*0}$
&$2.79^{+2.75}_{-1.97}$&$2.78^{+2.74}_{-1.96}$$^{+0.04}_{-0.06}$
&$1.94^{+2.52}_{-0.10}$&---&--- \\
&$\bar B_{s}\to \omega K^{*0}$
&$2.72^{+2.85}_{-1.83}$&$2.73^{+2.83}_{-1.81}$$^{+0.04}_{-0.05}$
&$2.18^{+0.33}_{-0.28}$ &---&---\\
\hline
$\phi_{\perp}+\pi$
&$\bar B_{s}\to \rho^{-}K^{*+}$
&$3.16^{+0.22}_{-0.24}$&$3.20^{+0.22}_{-0.24}$$^{+0.23}_{-0.22}$
&$3.07\pm0.08$ &---&---\\
&$\bar B_{s}\to \rho^{0}K^{*0}$
&$2.71^{+2.84}_{-1.87}$&$2.78^{+2.65}_{-1.96}$$^{+0.05}_{-0.10}$
&$1.99^{+2.53}_{-0.10}$&--- &---\\
&$\bar B_{s}\to \omega K^{*0}$
&$2.65^{+2.89}_{-1.75}$&$2.70^{+2.79}_{-1.82}$$^{+0.04}_{-0.08}$
&$2.23^{+0.32}_{-0.27}$ &---&---\\
\hline
$\Delta\phi_{\parallel}$
&$\bar B_{s}\to \rho^{-}K^{*+}$
&$0.13^{+0.01}_{-0.01}$&$0.12^{+0.01}_{-0.01}$$^{+0.06}_{-0.13}$
&$0.12^{+0.05}_{-0.05}$ &$0.06^{+0.02+0.11}_{-0.02-0.12}$&---\\
&$\bar B_{s}\to \rho^{0}K^{*0}$
&$-0.40^{+0.72}_{-0.28}$&$-0.33^{+0.73}_{-0.25}$$^{+0.33}_{-0.12}$
&$-0.32^{+2.74}_{-0.16}$&---&--- \\
&$\bar B_{s}\to \omega K^{*0}$
&$0.17^{+0.24}_{-0.61}$&$0.11^{+0.22}_{-0.61}$$^{+0.11}_{-0.34}$
&$0.31^{+0.31}_{-0.24}$ &---&---\\
\hline
$\Delta\phi_{\perp}$
&$\bar B_{s}\to \rho^{-}K^{*+}$
&$0.09^{+0.02}_{-0.02}$&$0.11^{+0.02}_{-0.02}$$^{+0.06}_{-0.14}$
&$0.12^{+0.04}_{-0.05}$ &---&---\\
&$\bar B_{s}\to \rho^{0}K^{*0}$
&$-0.41^{+0.85}_{-0.24}$&$-0.21^{+0.34}_{-0.30}$$^{+0.40}_{-0.14}$
&$-0.36^{+2.22}_{-0.16}$&---&--- \\
&$\bar B_{s}\to \omega K^{*0}$
&$0.20^{+0.75}_{-0.74}$&$-0.05^{+0.33}_{-0.20}$$^{+0.13}_{-0.38}$
&$0.36^{+0.31}_{-0.24}$ &---&---\\
\hline\hline
\end{tabular}
\end{center}
\end{table}

\begin{table}[htbp]
\begin{center}
\caption{\small Theoretical results for the $b\to s$ induced $\bar B_{s}\to K^{*-}K^{*+}$, $\rho^{0}\phi$ and $\omega \phi$ decays. The other captions are the same as in Table~\ref{Bs}.} \label{dels}
\vspace*{0.2cm}
\renewcommand{\arraystretch}{1.0}
\footnotesize \tabcolsep 0.04in
\begin{tabular}{llccccc}
\hline\hline
{Observable} &{Decay mode}&{Case I}&{Case II}&pQCD~\cite{Zou:2015iwa}  &QCDF~\cite{Beneke:2006hg}&QCDF~\cite{Cheng:2009mu}\\ \hline
${\cal B}[10^{-6}]$
&$\bar B_{s}\to K^{*-}K^{*+}$
&$20.0^{+1.0}_{-1.0}$&$9.7^{+0.9+6.9}_{-0.6-5.4}$&$5.4^{+2.7+1.8+0.3}_{-1.7-1.4-0.5}$ &$9.1^{+2.5+10.2}_{-2.2-5.9}$&$7.6^{+1.0+2.3}_{-1.0-1.8}$\\
&$\bar B_{s}\to \rho^{0}\phi$
&$0.41^{+0.11}_{-0.09}$&$0.41^{+0.11+0}_{-0.09-0}$&$0.23^{+0.15+0.03+0.01}_{-0.05-0.01-0.02}$&$0.40^{+0.12+0.27}_{-0.10-0.04}$&$0.18^{+0.01+0.09}_{-0.01-0.04}$ \\
&$\bar B_{s}\to \omega \phi$
&$0.25^{+0.34}_{-0.15}$&$0.25^{+0.34+0}_{-0.15-0}$&$0.17^{+0.10+0.05+0.00}_{-0.07-0.04-0.01}$ &$0.10^{+0.05+0.48}_{-0.03-0.12}$&$0.18^{+0.44+0.47}_{-0.12-0.04}$\\
\hline
$A_{CP}[\%]$
&$\bar B_{s}\to K^{*-}K^{*+}$
&$-34.4^{+4.3}_{-3.7}$&$20.7^{+5.8+8.4}_{-6.0-18.3}$&$8.8^{+2.5+0.5+0.0}_{-8.9-2.9-0.2}$ &$2^{+0+40}_{-0-15}$&$21^{+1+2}_{-2-4}$\\
&$\bar B_{s}\to \rho^{0}\phi$
&$24.5^{+25.1}_{-20.9}$&$24.5^{+25.1+0}_{-20.9-0}$&$-4.3^{+0.6+0.6+1.2}_{-0.5-0.5-1.0}$&$19^{+5+56}_{-5-67}$&$83^{+1+10}_{-0-36}$ \\
&$\bar B_{s}\to \omega \phi$
&$-14.5^{+18.9}_{-18.7}$&$-14.5^{+18.9+0}_{-18.7-0}$&$28.0^{+1.3+0.5+3.4}_{-3.2-2.3-5.1}$ &$8^{+3+102}_{-3-56}$&$-8^{+3+20}_{-1-15}$\\
\hline
$A_{CP}^0[\%]$
&$\bar B_{s}\to K^{*-}K^{*+}$
&$-15.5^{+3.2}_{-2.9}$&$40.5^{+15.3+41.9}_{-13.6-37.3}$&$45.4^{+19.0}_{-23.4}$ &$11^{+3+7}_{-3-17}$&---\\
&$\bar B_{s}\to \rho^{0}\phi$
&$-4.5^{+24.6}_{-12.9}$&$-4.5^{+24.6+0}_{-12.9-0}$&$3.27^{+1.07}_{-1.19}$&$11^{+4+10}_{-3-8}$&---\\
&$\bar B_{s}\to \omega \phi$
&$0.2^{+3.7}_{-2.5}$&$0.2^{+3.7+0}_{-2.5-0}$&$-2.24^{+6.67}_{-5.45}$ &---&---\\
\hline
$A_{CP}^{\bot}[\%]$
&$\bar B_{s}\to K^{*-}K^{*+}$
&$37.1^{+4.4}_{-4.4}$&$-31.6^{+5.5+21.1}_{-5.3-12.8}$&$-32.9^{+5.6}_{-4.0}$ &---&---\\
&$\bar B_{s}\to \rho^{0}\phi$
&$71.7^{+16.3}_{-16.7}$&$71.7^{+16.3+0}_{-16.7-0}$&$-32.8^{+7.4}_{-5.8}$&---&--- \\
&$\bar B_{s}\to \omega \phi$
&$-1.0^{+11.4}_{-6.5}$&$-1.0^{+11.4+0}_{-6.5-0}$&$4.38^{+17.52}_{-15.93}$ &---&---\\
\hline
$f_L[\%]$
&$\bar B_{s}\to K^{*-}K^{*+}$
&$64.4^{+1.2}_{-2.7}$&$33.9^{+4.7+9.9}_{-4.0-14.2}$&$42^{+13+3+5}_{-9-3-6}$ &$67^{+4+31}_{-5-26}$&$52^{+3+20}_{-5-21}$\\
&$\bar B_{s}\to \rho^{0}\phi$
&$93.8^{+1.6}_{-12.1}$&$93.8^{+1.6+0}_{-12.1-0}$&$86^{+1+1+0}_{-1-1-0}$&$81^{+3+9}_{-4-12}$&$88^{+1+2}_{-0-18}$ \\
&$\bar B_{s}\to \omega \phi$
&$55.1^{+32.7}_{-16.3}$&$55.1^{+32.7+0}_{-16.3-0}$&$69^{+8+8+2}_{-9-9-2}$ &---&$95^{+1+0}_{-2-42}$\\
\hline
$f_{\bot}[\%]$
&$\bar B_{s}\to K^{*-}K^{*+}$
&$16.1^{+0.9}_{-0.8}$&$40.6^{+3.9+5.6}_{-3.6-12.9}$&$27.7^{+5.2}_{-7.0}$ &---&---\\
&$\bar B_{s}\to \rho^{0}\phi$
&$2.8^{+5.7}_{-0.8}$&$2.8^{+5.7+0}_{-0.8-0}$&$8.89^{+0.80}_{-1.06}$&---&--- \\
&$\bar B_{s}\to \omega \phi$
&$23.0^{+8.4}_{-16.9}$&$23.0^{+8.4+0}_{-16.9-0}$&$16.1^{+7.3}_{-5.8}$ &---&---\\
\hline
$\phi_{\parallel}+\pi$
&$\bar B_{s}\to K^{*-}K^{*+}$
&$2.86^{+0.10}_{-0.09}$&$1.94^{+0.10+0.59}_{-0.11-0.54}$&$3.53^{+0.33}_{-0.25}$ &$2.84^{+0.02+1.00}_{-0.03-0.61}$&---\\
&$\bar B_{s}\to \rho^{0}\phi$
&$3.01^{+0.22}_{-0.33}$&$3.01^{+0.22+0}_{-0.33-0}$&$3.11^{+0.10}_{-0.09}$&$1.60^{+0.05+0.10}_{-0.06-0.15}$&---\\
&$\bar B_{s}\to \omega \phi$
&$2.90^{+0.24}_{-0.30}$&$2.90^{+0.24+0}_{-0.30-0}$&$3.38^{+0.20}_{-0.17}$ &$2.00^{+0+0.44}_{-0-0.87}$&---\\
\hline
$\phi_{\perp}+\pi$
&$\bar B_{s}\to K^{*-}K^{*+}$
&$2.76^{+0.10}_{-0.11}$&$1.78^{+0.13+0.13}_{-0.13-0.12}$&$3.54^{+0.36}_{-0.24}$ &---&---\\
&$\bar B_{s}\to \rho^{0}\phi$
&$2.95^{+0.31}_{-0.34}$&$2.95^{+0.31+0}_{-0.34-0}$&$3.29\pm0.09$&--- &---\\
&$\bar B_{s}\to \omega \phi$
&$2.92^{+0.19}_{-0.14}$&$2.92^{+0.19+0}_{-0.14-0}$&$3.35^{+0.30}_{-0.23}$ &---&---\\
\hline
$\Delta\phi_{\parallel}$
&$\bar B_{s}\to K^{*-}K^{*+}$
&$-0.11^{+0.04}_{-0.04}$&$0.49^{+0.14+0.43}_{-0.13-0.20}$&$0.94^{+0.11}_{-0.14}$ &$-0.15^{+0.03+0.92}_{-0.03-0.17}$&---\\
&$\bar B_{s}\to \rho^{0}\phi$
&$0.41^{+0.32}_{-0.19}$&$0.41^{+0.32+0}_{-0.19-0}$&$-0.44\pm0.10$&$-0.13^{+0.03+0.17}_{-0.03-0.13}$&---\\
&$\bar B_{s}\to \omega \phi$
&$0.24^{+0.17}_{-0.09}$&$0.24^{+0.17+0}_{-0.09-0}$&$-0.37\pm0.12$ &$0.11^{+0.03+0.70}_{-0.03-0.63}$&---\\
\hline
$\Delta\phi_{\perp}$
&$\bar B_{s}\to K^{*-}K^{*+}$
&$-0.01^{+0.03}_{-0.04}$&$0.50^{+0.09+0.37}_{-0.10-0.20}$&$0.93^{+0.11}_{-0.14}$ &---&---\\
&$\bar B_{s}\to \rho^{0}\phi$
&$0.51^{+0.34}_{-0.37}$&$0.51^{+0.34+0}_{-0.37-0}$&$-0.64^{+0.11}_{-0.10}$&--- &---\\
&$\bar B_{s}\to \omega \phi$
&$0.26^{+0.18}_{-0.10}$&$0.26^{+0.18+0}_{-0.10-0}$&$-0.33^{+0.16}_{-0.19}$ &---&---\\
\hline\hline
\end{tabular}
\end{center}
\end{table}

\begin{table}[htbp]
\begin{center}
\caption{\small Theoretical results for the pure annihilation $\bar B_{s}\to \rho\rho$, $\rho\omega$ and $\omega\omega$ decays. Our results $A_{CP}^0=0$, $\Delta\phi_{\perp}=0$, $\Delta\phi_{\parallel}=0$ are in agreement with previous ones and are, therefore, not listed here. The other captions are the same as in Table~\ref{Bs}.} \label{anni}
\vspace*{0.2cm}
\renewcommand{\arraystretch}{1.0}
\footnotesize \tabcolsep 0.04in
\begin{tabular}{llccccc}
\hline\hline
{Observable} &{Decay mode}&{Case I}&{Case II}&pQCD~\cite{Zou:2015iwa}  &QCDF~\cite{Beneke:2006hg} &QCDF~\cite{Cheng:2009mu}\\ \hline
${\cal B}[10^{-6}]$
&$\bar B_{s}\to \rho^{+}\rho^{-}$
&$25.6^{+1.3}_{-1.3}$&$10.4^{+0.2+6.7}_{-0.3-4.9}$&$1.5^{+0.7+0.2+0.0}_{-0.6-0.2-0.1}$&$0.34^{+0.03+0.60}_{-0.03-0.38}$&$0.68^{+0.04+0.73}_{-0.04-0.53}$\\
&$\bar B_{s}\to \rho^{0}\rho^{0}$
&$15.2^{+4.2}_{-3.8}$&$5.2^{+0.1+3.4}_{-0.1-2.5}$ &$0.74^{+0.39+0.22+0.00}_{-0.24-0.14-0.00}$&$0.17^{+0.01+0.30}_{-0.01-0.19}$&$0.34^{+0.02+0.36}_{-0.02-0.26}$\\
&$\bar B_{s}\to \rho^{0}\omega$
&$0.05^{+0.01}_{-0.01}$&$0.02^{+0.00+0.01}_{-0.00-0.01}$&$0.009^{+0.003+0.001+0.000}_{-0.003-0.002-0.001}$&$<0.01$ &$0.004^{+0.0+0.005}_{-0.0-0.003}$\\
&$\bar B_{s}\to \omega \omega$
&$9.6^{+1.6}_{-1.4}$&$3.9^{+0.7+2.5}_{-0.6-1.9}$&$0.40^{+0.16+0.10+0.00}_{-0.18-0.10-0.01}$ &$0.11^{+0.01+0.20}_{-0.01-0.12}$&$0.19^{+0.02+0.21}_{-0.02-0.15}$\\
\hline
$A_{CP}[\%]$
&$\bar B_{s}\to \rho^{+}\rho^{-}$
&$0\pm0$&$0\pm0$&$-2.9^{+0.7+1.5+0.2}_{-1.1-1.3-0.2}$&---&$0$\\
&$\bar B_{s}\to \rho^{0}\rho^{0}$
&$0\pm0$&$0\pm0$&$-2.9^{+0.7+1.5+0.2}_{-1.1-1.3-0.2}$&---&$0$\\
&$\bar B_{s}\to \rho^{0}\omega$
&$0\pm0$&$0\pm0$&$11.1^{+1.0+1.9+1.2}_{-1.5-4.4-1.4}$&---&$0$\\
&$\bar B_{s}\to \omega \omega$
&$0\pm0$&$0\pm0$&$-3.3^{+0.8+1.5+0.5}_{-1.0-1.4-0.2}$&---&$0$\\
\hline
$A_{CP}^{\bot}[\%]$
&$\bar B_{s}\to \rho^{+}\rho^{-}$
&$0\pm0$&$0\pm0$&$30.5^{+15.0}_{-16.3}$ &---&---\\
&$\bar B_{s}\to \rho^{0}\rho^{0}$
&$0\pm0$&$0\pm0$&$30.5^{+15.0}_{-16.3}$ &---&---\\
&$\bar B_{s}\to \rho^{0}\omega$
&$0\pm0$&$0\pm0$&$27.9^{+9.3}_{-9.9}$&---&---\\
&$\bar B_{s}\to \omega \omega$
&$0\pm0$&$0\pm0$&$30.8^{+14.0}_{-15.3}$ &---&---\\
\hline
$f_L[\%]$
&$\bar B_{s}\to \rho^{+}\rho^{-}$
&$83\pm1$&$55^{+1+13}_{-1-9}$&$\sim100$ &---&$\sim100$\\
&$\bar B_{s}\to \rho^{0}\rho^{0}$
&$83\pm1$&$55^{+1+13}_{-1-9}$&$\sim100$ &---&$\sim100$\\
&$\bar B_{s}\to \rho^{0}\omega$
&$83\pm1$&$55^{+1+13}_{-1-9}$&$\sim100$ &---&$\sim100$\\
&$\bar B_{s}\to \omega \omega$
&$83\pm2$&$54^{+1+13}_{-1-9}$&$\sim100$ &---&$\sim100$\\
\hline
$f_\perp[\%]$
&$\bar B_{s}\to \rho^{+}\rho^{-}$
&$8\pm1$&$25^{+0+5}_{-0-8}$&$\sim0.0$ &---&---\\
&$\bar B_{s}\to \rho^{0}\rho^{0}$
&$8\pm1$&$25^{+0+5}_{-0-8}$&$\sim0.0$ &---&---\\
&$\bar B_{s}\to \rho^{0}\omega$
&$8\pm1$&$25^{+1+5}_{-1-8}$&$\sim0.0$&---&---\\
&$\bar B_{s}\to \omega \omega$
&$8\pm1$&$25^{+1+5}_{-1-7}$&$\sim0.0$ &---&---\\
\hline
$\phi_{\parallel}+\pi$
&$\bar B_{s}\to \rho^{+}\rho^{-}$
&$2.60\pm0.20$&$1.79^{+0.10+0.24}_{-0.10-0.22}$&$3.40\pm0.04$ &---&---\\
&$\bar B_{s}\to \rho^{0}\rho^{0}$
&$2.61\pm0.31$&$1.79^{+0.10+0.24}_{-0.10-0.22}$&$3.40\pm0.04$ &---&---\\
&$\bar B_{s}\to \rho^{0}\omega$
&$2.61\pm0.30$&$1.80^{+0.12+0.24}_{-0.12-0.22}$&$3.48\pm0.04$&---&---\\
&$\bar B_{s}\to \omega \omega$
&$2.60\pm0.50$&$1.80^{+0.12+0.24}_{-0.12-0.22}$&$3.40\pm0.04$ &---&---\\
\hline
$\phi_{\perp}+\pi$
&$\bar B_{s}\to \rho^{+}\rho^{-}$
&$2.59\pm0.21$&$1.68^{+0.10+0.23}_{-0.10-0.23}$&$3.27^{+0.16}_{-0.15}$ &---&---\\
&$\bar B_{s}\to \rho^{0}\rho^{0}$
&$2.59\pm0.32$&$1.68^{+0.12+0.23}_{-0.12-0.23}$&$3.27^{+0.16}_{-0.15}$ &---&---\\
&$\bar B_{s}\to \rho^{0}\omega$
&$2.59\pm0.31$&$1.68^{+0.10+0.23}_{-0.10-0.23}$&$2.63^{+0.18}_{-0.22}$&---&---\\
&$\bar B_{s}\to \omega \omega$
&$2.58\pm0.50$&$1.70^{+0.12+0.23}_{-0.12-0.23}$&$3.27^{+0.16}_{-0.11}$ &---&---\\
\hline\hline
\end{tabular}
\end{center}
\end{table}

Using the fitted results given by Eqs.~\eqref{eq:bff1},~\eqref{eq:bfi1}~(case I) and \eqref{scasIIi}, \eqref{scasIIf}~(case II) for $(\rho_{A}^{i,f}, \phi_{A}^{i,f})$, Eq.~\eqref{eq:datarhophi} for $(\rho_{H}, \phi_{H})$, and the other input parameters listed in Table~\ref{ppvalue},
we now present in Tables~\ref{Bs}, \ref{deld}, \ref{dels} and \ref{anni} our updated theoretical results
~(posterior predictions\footnote{Our updated theoretical results in Tables~\ref{Bs}, \ref{deld}, \ref{dels} and \ref{anni} should be treated as posterior predictions since they are based on the current data and the model for the end-point regularization of HSS and WA contributions in QCDF. }) 
for the branching ratios, CP asymmetries, polarization fractions and relative phases in $\bar B_{s}\to \rho K^{*}$, $\omega K^{*}$, $\phi K^{*}$, $\bar{K}^{*} K^{*}$, $\phi \phi$, $\rho \phi$, $\omega \phi$, $\rho \rho$, $\rho \omega$, $\omega \omega$ decays, where the previous predictions in the QCDF~\cite{Beneke:2006hg,Cheng:2009mu} and pQCD~\cite{Zou:2015iwa} approaches are also listed for comparison. The first uncertainty for the results of cases I and II in these tables is caused by the input parameters listed in Table~\ref{ppvalue} and $\phi_{H}$ given by Eq.~\eqref{eq:datarhophi}; the second uncertainty for the results of case II corresponds to the uncertainties of $(\rho_{A}^{i,f}, \phi_{A}^{i,f})$ given by Eqs.  \eqref{scasIIi} and  \eqref{scasIIf}.

One can find from these tables that most of our results for the observables are in agreement with the current experimental data including the abnormal $f_{L,\bot}(\bar B_{s}\to K^{*0}\bar K^{*0})$; the only exception is for ${\cal B}(\bar B_{s}\to \phi K^{*0})$. A detailed discussion has been presented in the last subsections. More theoretical and experimental efforts are needed to confirm or refute this possible puzzle.

Our results are also generally in consistence with the previous theoretical predictions in QCDF~\cite{Beneke:2006hg,Cheng:2009mu} and pQCD~\cite{Zou:2015iwa} within the theoretical uncertainties. The most obvious differences are the results for the pure annihilation $B_s$ decays, which can be seen from Table~\ref{anni}. Our results for the branching ratios of $\bar B_{s}\to \rho\rho$, $\rho\omega$ and $\omega \omega$ decays are about one order larger than the previous predictions; moreover, our results, $f_{L,\bot}\sim 55\%\,,25\%$, are also obviously different from $f_{L,\bot}\sim 100\%\,,0\%$~\cite{Beneke:2006hg,Zou:2015iwa}. These differences in fact can be easily understood from the following: (i) The best-fit value of $\rho^i_A$  is very large in order to fit the abnormal $f_{L,\bot}(\bar B_{s}\to K^{*0}\bar K^{*0})\sim 20\%\,,58\%$ in case II as discussed above; it results in sizable nonfactorizable annihilation contributions. (ii) The pure annihilation decays,  $\bar B_{s}\to \omega \phi$, $\rho \rho$, $\rho \omega$, $\omega \omega$, are only relevant to the nonfactorizable annihilation amplitudes. Therefore, it can be briefly concluded that, if one requires the WA corrections to account for the abnormal $f_{L,\bot}(\bar B_{s}\to K^{*0}\bar K^{*0})$ measured by the LHCb collaboration, the large branching ratios and transverse polarization fractions of $\bar B_{s}\to \rho\rho$, $\rho\omega$ and $\omega \omega$ decays will be expected accordingly.  Interestingly, the large nonfactorizable annihilation contributions have been observed in the pure annihilation $B_s\to\pi^+\pi^-$  decay~\cite{Zhu:2011mm,Wang:2013fya,Bobeth:2014rra,CDFanni,Duh:2012ie,LHCbanni}.

Finally, we would like to point out that the allowed spaces for the end-point parameters are still very large, especially in case I; our results are only based on the best-fit points, and the other allowed spaces  for $(\rho_{A}^{i,f}, \phi_{A}^{i,f})$  shown in Figs. \ref{c1} and \ref{c1pp} are not taken into account here; moreover, it is also not clear whether the annihilation corrections should account for the abnormal $f_{L,\bot}(\bar B_{s}\to K^{*0}\bar K^{*0})$. More data on $B_s\to VV$ decays are needed for a definite conclusion.  The pure annihilation decays mentioned above, except for  $\bar B_{s}\to \rho\omega$, having branching ratios $\gtrsim {\cal O}(10^{-7})$ are in the scope of the LHCb and Belle-II experiments. Hence, a much clearer picture of the WA contributions in charmless $B_s\to VV$ decays is expected to be obtained from these dedicated heavy-flavor experiments in the near future.

\section{Conclusion}

In summary, we have studied the HSS and WA contributions in charmless $B_s\to VV$ decays. In order to probe their strength and possible strong phase, we have performed $\chi^2$-analyses for the end-point parameters under the constraints from the measured $\bar B_{s}\to$$\rho^0\phi$, $\phi K^{*0}$, $\phi \phi$ and $K^{*0}\bar K^{*0}$ decays. It is found that the end-point parameters in the factorizable and nonfactorizable annihilation topologies are non-universal at 68\% C.L. due to the constraint from $\bar B_{s}\to K^{*0}\bar K^{*0}$ decay; this further confirms the findings in the previous work. Moreover, the abnormal polarization fractions $f_{L,\bot}(\bar B_{s}\to K^{*0}\bar K^{*0})=(20.1\pm7.0)\%\,,(58.4\pm8.5)\%$ measured by the LHCb collaboration can be reconciled through the weak annihilation corrections with a large $\rho_A^i$.
However, the ${\cal B}(\bar B_{s}\to\phi K^{*0})$ exhibits a significant tension between the data and theoretical results, which dominates the contributions to $\chi_{\rm min}^2$ in the fits.
Using the best-fit end-point parameters, we have also updated the theoretical results for the charmless $B_s\to VV$ decays within the  framework of QCDF. It is found that the large branching fractions and transverse polarization fractions for the pure annihilation decays are possible if we require the WA contributions to account for the abnormal $f_{L,\bot}(\bar B_{s}\to K^{*0}\bar K^{*0})$. Our results and findings will be further tested by the LHCb and Belle-II experiments in the near future.

\section*{Acknowledgements}
We thank Dr. Xingbo Yuan at NCTS for helpful discussions. The work is supported by the National Natural Science Foundation of China under contract Nos.  11475055, 11675061, 11435003  and U1632109. Q.~Chang is also supported by the Foundation for the Author of National Excellent Doctoral Dissertation of P.~R.~China (Grant No. 201317), the Program for Science and Technology Innovation Talents in Universities of Henan Province (Grant No. 14HASTIT036), the Excellent Youth Foundation of HNNU and the CSC.

 \end{document}